\documentclass[emulateapj]{aastex62} %

\received{} %
\revised{} %
\accepted{} %
\submitjournal{ApJ} %

\shorttitle{Filamentary Accretion Flows in IRDC G14.2} %
\shortauthors{Chen et al.} %

\begin{document} %

\title{Filamentary Accretion Flows in the Infrared Dark Cloud G14.225$-$0.506 Revealed by ALMA} %

\correspondingauthor{Huei-Ru Vivien Chen} %
\email{hchen@phys.nthu.edu.tw} %

\author[0000-0002-9774-1846]{Huei-Ru Vivien Chen} %
\affil{Institute of Astronomy and Department of Physics, National Tsing Hua University, Hsinchu 30013, Taiwan} %

\author[0000-0003-2384-6589]{Qizhou Zhang} %
\affiliation{Center for Astrophysics $|$ Harvard \& Smithsonian, 60 Garden Street, Cambridge, MA 02318, USA} %

\author[0000-0002-9154-2440]{M.~C.~H. Wright} %
\affiliation{Department of Astronomy, University of California, Berkeley, CA 94720, USA} %

\author[0000-0002-2189-6278]{Gemma Busquet} %
\affiliation{Institut de Ci\`encies de l'Espai (ICE, CSIC), Can Magrans, s/n, E-08193 Cerdanyola del Vall\`es, Catalonia \\
and Institut d'Estudis Espacials de Catalunya (IEEC), E-08034, Barcelona, Catalonia
} %

\author[0000-0001-9299-5479]{Yuxin Lin} %
\affiliation{Max-Planck-Institut f\"ur Radioastronomie, D-53121 Bonn, Germany} %

\author[0000-0003-2300-2626]{Hauyu Baobab Liu} %
\affiliation{European Southern Observatory (ESO), Karl-Schwarzschild-Str. 2, D-85748 Garching, Germany} %
\affiliation{(moving to) Academia Sinica Institute of Astronomy and Astrophysics, P.O. Box 23-141, Taipei 10617, Taiwan} %

\author[0000-0002-8250-6827]{F. A. Olguin} %
\affil{Institute of Astronomy, National Tsing Hua University, Hsinchu 30013, Taiwan} %

\author[0000-0002-7125-7685]{Patricio Sanhueza} %
\affiliation{National Astronomical Observatory of Japan, National Institutes of Natural Sciences, 2-21-1 Osawa, Mitaka, Tokyo 181-8588, Japan} %

\author[0000-0001-5431-2294]{Fumitaka Nakamura} %
\affiliation{National Astronomical Observatory of Japan, National Institutes of Natural Sciences, 2-21-1 Osawa, Mitaka, Tokyo 181-8588, Japan} %

\author[0000-0002-9569-9234]{Aina Palau} %
\affiliation{Instituto de Radioastronom\'ia y Astrof\'isica, Universidad Nacional Aut\'onoma de M\'exico, P.O. Box 3-72, 58090 Morelia, Michoac\'an, M\'exico} %

\author{Satoshi Ohashi} %
\affiliation{RIKEN, 2-1, Hirosawa, Wako-shi, Saitama 351-0198, Japan} %
\affiliation{National Astronomical Observatory of Japan, National Institutes of Natural Sciences, 2-21-1 Osawa, Mitaka, Tokyo 181-8588, Japan} %

\author[0000-0002-8149-8546]{Ken'ichi Tatematsu} %
\affiliation{National Astronomical Observatory of Japan, National Institutes of Natural Sciences, 2-21-1 Osawa, Mitaka, Tokyo 181-8588, Japan} %
\affiliation{Department of Astronomical Science, SOKENDAI (The Graduate University for Advanced Studies), 2-21-1 Osawa, Mitaka, Tokyo 181-8588, Japan} %

\author[0000-0002-7026-6782]{Li-Wen Liao} %
\affil{Institute of Astronomy, National Tsing Hua University, Hsinchu 30013, Taiwan} %




\begin{abstract}
Filaments are ubiquitous structures in molecular clouds and play an important role in the mass assembly of stars.  
We present results of dynamical stability analyses for filaments in the infrared dark cloud G14.225$-$0.506, where a delayed onset of massive star formation was reported in the two hubs at the convergence of multiple filaments of parsec length.   
Full-synthesis imaging is performed with the  Atacama Large Millimeter/submillimeter Array (ALMA) to map the $\mathrm{N_2H^+} \; (1-0)$ emission in two hub-filament systems with a spatial resolution of $\sim 0.034 \; \mathrm{pc}$.    
Kinematics are derived from sophisticated spectral fitting algorithm that accounts for line blending, large optical depth, and multiple velocity components.  
We identify five velocity coherent filaments and derive their velocity gradients with principal component analysis.  
The mass accretion rates along the filaments are up to $10^{-4} \; \mathrm{M_\odot \, \mathrm{yr^{-1}}}$ and are significant enough to affect the hub dynamics within one free-fall time ($\sim 10^5 \; \mathrm{yr}$). 
The $\mathrm{N_2H^+}$ filaments are in equilibrium with virial parameter $\alpha_\mathrm{vir} \sim 1.2$.  
We compare $\alpha_\mathrm{vir}$ measured in the $\mathrm{N_2H^+}$ filaments, $\mathrm{NH_3}$ filaments, $870 \; \mu\mathrm{m}$ dense clumps, and $3 \; \mathrm{mm}$ dense cores. 
The decreasing trend in $\alpha_\mathrm{vir}$ with decreasing spatial scales persists, suggesting an increasingly important role of gravity at small scales. 
Meanwhile, $\alpha_\mathrm{vir}$ also decreases with decreasing non-thermal motions.  
In combination with the absence of high-mass protostars and massive cores, our results are consistent with the global hierarchical collapse scenario.  
\end{abstract}



\section{Introduction} \label{sec:intro} %
How accretion proceeds around young star clusters affects the mass growth of protostars and is critical to the understanding of the origin of the initial mass function (IMF).    
The lack of observational characterization of young cluster-forming regions precludes a unified theoretical scenario to explain star formation across several orders of magnitude in mass and scales.  
Recent {\it Herschel} observations reveal that parsec-scale filaments are prevalent in molecular clouds \citep[e.g.][]{Andre:2010ka,Molinari:2010hm,Arzoumanian:2011ho,Arzoumanian:2013gp,Palmeirim:2013da}.  
Young stellar groups are often found in dense clumps of column density exceeding $10^{22} \; \mathrm{cm^{-2}}$ at the convergence of multiple filaments of parsec length, namely ``hub-filament systems'' \citep{Myers:2009if,Liu:2012ex,Liu:2015nq,Lu:2018bn,Peretto:2013kt,Peretto:2014cx,Williams:2018jn}.  
Based on core accretion scenarios \citep[e.g.][]{Shu:1977ef,McKee:2003gx}, theoretical models have gradually incorporated accretion from the surrounding clumps \citep[e.g.][]{Bate:2005cr,Wang:2010ib,Myers:2011bg,Myers:2013jf}.  
Cores embedded in denser clumps benefit by accretion from the filamentary environment so as to prolong the accretion time for growing massive stars \citep{Myers:2009fv}. 
Meanwhile, numerical simulations of colliding flows and collapsing turbulent clumps grow massive protostars from low-mass stellar seeds by feeding gas along the dense filamentary streams converging toward the $0.1 \; \mathrm{pc}$-size hubs with detectable velocity gradients along the filaments  \citep[e.g.][]{Wang:2010ib,Gomez:2014bu,Smith:2016dn}.
Although filaments are expected in colliding flows, their origin and internal structures remain debatable \citep{Smith:2016dn,Moeckel:2015cj,Clarke:2017cm}.  
To date, only a few spectral line observations have been conducted to trace the hypothesized accretion flows along filaments, presumably towards the center of gravity, where proto-clusters are located \citep[e.g.][]{Kirk:2013gq,Peretto:2013kt,Lu:2018bn,Liu:2012ex}.

At a distance of $1.98_{-0.12}^{+0.13} \; \mathrm{kpc}$ \citep{Xu:2011da},  the infrared dark cloud (IRDC) G14.225$-$0.506 (hereafter G14.2) is part of the remarkable IRDC complex, M17~SWex \citep[Fig.~\ref{fig:mosaic}a;][]{Povich:2010iv}, which was first discovered by \citet{Elmegreen:1976jo} in the CO map as a large ($67 \; \mathrm{pc} \times 17 \; \mathrm{pc}$) and massive ($\sim 3 \times 10^5 \; M_\sun$) molecular cloud complex extended parallel to the Galactic plane southwest of the well known giant H~{\footnotesize II} region M17.  
In the most extincted part of M17~SWex, the $\mathrm{NH_3} \; (1,1)$ emission reveals a network of filaments associated with two warmer ($T_\mathrm{rot} \sim 15 \; \mathrm{K}$) hubs, hub-N and hub-S, at the convergence of multiple cold, velocity coherent filaments ($\sim 10 \; \mathrm{K}$) of parsec lengths at distinct velocities \citep{Busquet:2013ko}.  
The velocity dispersion in hubs is a factor of $\sim 2$ broader than in filaments.  
The larger velocity dispersion in the hubs may be due to higher temperature, star formation activities, colliding filaments \citep{Wang:2010ib}, or longitudinally collapsing filaments \citep{Peretto:2014cx}.  
Analyses of young stellar objects (YSOs) based on near- and mid-infrared {\it Spitzer} photometry data together with the {\it Chandra} X-ray census (for diskless YSOs) reveal a rich population of intermediate-mass YSOs without commensurate, simultaneous massive star formation \citep{Povich:2010iv,Povich:2016jm}.  
Such conspicuous deficit of O-type massive protostars implies that the IRDC~G14.2 may be either an example of a distributed star formation mode with OB clusters dominated by intermediate-mass stars or its massive hubs/cores are still in the process of accreting ambient material to nurture massive protostars.  
In the latter case, the high-mass tail in the protostar mass function (PMF) will arise later in time \citep{Bonnell:2006ee,Myers:2009fv,VazquezSemadeni:2017gq}.   
\citet{Ohashi:2016iz} have performed a dense core survey in IRDC~G14.2 using the $3 \; \mathrm{mm}$ continuum emission (angular resolutions of $\sim 3\arcsec \times 2\arcsec$ and a sensitivity of $0.28 \; M_\sun$) in two mosaic fields covering the two hubs and their associated networks of filaments (see Fig.~\ref{fig:mosaic}b).  
The maximum mass of the prestellar or protostellar cores ($\lesssim 22 \; M_\sun$) suggests a scenario of forming high-mass stars in prestellar cores by accreting significant amount of gas from the surroundings or prolonging the accretion from protostellar cores to intermediate-mass YSOs. 
The hubs contain more mass and have potential to nurture massive stars.  
The total gas mass estimated by \citet{Ohashi:2016iz} is  $1400 \; M_\odot$ in the hub-N and $960 \; M_\odot$ in the hub-S.   
Assuming a star formation efficiency of 30\% and the initial mass function (IMF) from \citet{Kroupa:2001ki}, we estimate the expected maximum stellar mass to be $27 \; M_\odot$ in the hub-N and $21 \; M_\odot$ in the hub-S \citep[apply Equation~(2) in][]{Sanhueza:2017bd}.  
If so, IRDC~G14.2 is one of the most ideal systems to characterize the initial conditions of massive star formation.  

To trace quiescent gas kinematics in G14.2, we choose to map emission of the molecular ion, $\mathrm{N_2H^+}$, which has a much higher critical density, $n_\mathrm{crit} \sim 10^5 \; \mathrm{cm^{-3}}$, and a lower upper-level energy, $E_\mathrm{up} = 4.5 \; \mathrm{K}$, than the $\mathrm{NH_3} \; (1,1)$ line with $n_\mathrm{crit} \sim 10^3 \; \mathrm{cm^{-3}}$ and $E_\mathrm{up} = 23.4 \; \mathrm{K}$ \citep{Shirley:2015et}.  
The multiple-spin coupling induced by the two nitrogen nuclei in the molecular ion, $\mathrm{N_2H^+}$, gives rise to splitting of the $J= 1-0$ line into seven closely spaced hyperfine components \citep{Green:1974gc}, which are often observed in high-mass star-forming regions and IRDCs as a triplet of lines \citep{Caselli:1995hl,Shirley:2005hy,Sanhueza:2012iy}. 
Only one isolated component, $JF_1F = 101 \rightarrow 012$, is well separated from the other six hyperfine components and can be used to directly trace gas kinematics without the need to fit all the components.  
This isolated component, however, is fairly weak and has a relative intensity of merely $1/9 \simeq 0.11$ of the total intensity \citep{Mangum:2015ch}.  

In this paper, we present full-synthesis images of the $\mathrm{N_2H^+} \; (1-0)$ line obtained with the Atacama Large Millimeter/submillimeter Array (ALMA) in the two mosaic fields. 
The continuum counterpart of the data were previously reported by \citet{Ohashi:2016iz}.   
Details of $\mathrm{N_2H^+}$ line observations are described in Section~\ref{sec:obs}.  
The morphology of dense molecular gas and its relation with the embedded YSOs are discussed in Section~\ref{sec:results}.  
Identification and kinematics analyses of filaments are described in Section~\ref{sec:analyses}. 
We then discuss the main results in Section~\ref{sec:discussion}, and conclude in Section~\ref{sec:conclusion}.

\section{Observations} \label{sec:obs} %
\begin{figure}[h!] %
\epsscale{1.0} %
\plotone{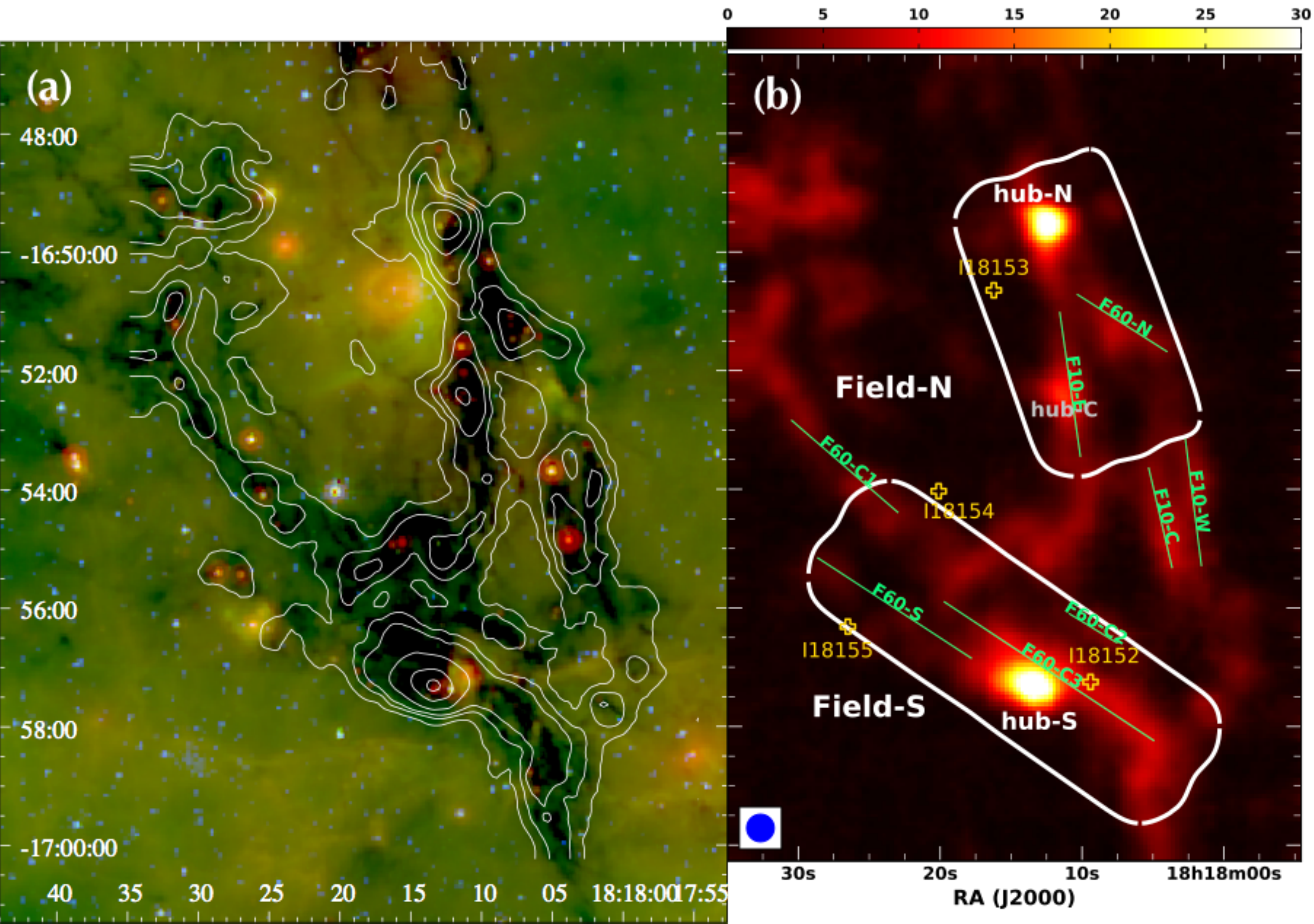} %
\caption{(a) Archival {\it Spitzer} $3.6/8/24 \mu\mathrm{m}$ (blue/green/red) three-color composite image of G14.225$-$0.506 overlaid with the $\mathrm{N_2H^+} \; (1-0)$ integrated intensity map (contours) observed with the IRAM 30-m Telescope (Busquet et~al. 2019, in prep.).  Contour levels are $(4,8,12,24,48,72)\times \sigma$, where the rms noise level is $\sigma = 0.05 \; \mathrm{K \, km \, s^{-1}}$. (b) ALMA mosaic fields, Field-N and Field-S (white boxes) overlaid on the same $\mathrm{N_2H^+} \; (1-0)$ integrated intensity map (color scale) as in (a).  The two massive star-forming hubs, hub-N and hub-S, are labeled.  Yellow open crosses indicate the IRAS sources in the field of view. Green lines with labels indicate the positions of the previously identified $\mathrm{NH_3}$ filaments \citep{Busquet:2013ko}. \label{fig:mosaic}} %
\end{figure}

The IRDC G14.2 was observed with the ALMA 12-m Array on 2015 April 25 in the C34-2/1 configuration with a total of 37 antennas and with ACA (7-m Array antennas) on 2015 April 30 and May 4, 2016 May 15 and June 4 with a total of 10 antennas (Cycle 2  and 3 programs, Project ID: 2013.1.00312.S and 2015.1.00418.S; PI: Vivien Chen). 
Observations with the total power (TP) array were conducted from 2016 May 14 to May 20 in multiple sessions.  
The total number of the 12-m array pointings is 57 in Field-N and 67 in Field-S (Fig.~\ref{fig:mosaic}b).  
The duration of all the 12-m array observation, including time for calibration, is roughly 1.7~hr.  
All observations employed the Band~3 receivers with an instrumental spectral resolution of $31 \; \mathrm{kHz}$ 
centered at the rest frequency of $93.1738 \; \mathrm{GHz}$ for the $\mathrm{N_2H^+} \; (1-0)$ transition.   
The system temperatures ranged from 60 to $90 \; \mathrm{K}$. 
The projected baselines, including both 7-m and 12-m arrays, ranged from $2.6$ to $107 \; \mathrm{k \lambda}$, equivalent to $0\farcs85$ to $35\arcsec$.  
The quasars J1733$-$1304 and J1924$-$2914 were observed for bandpass, phase, and amplitude calibration. 
Flux calibration was performed using Neptune and Ceres. 
The uncertainty of absolute flux calibration is 5\% in Band~3 according to ALMA Cycle~2 Technical Handbook.
The reduction and calibration of the data were done with CASA version 4.3.1, 4.5.3, and 4.7.0 \citep{McMullin:2007tj} using the standard procedures, and the data were delivered from the East Asian ALMA Regional Center. 
The visibility data were then exported in FITS format to the MIRIAD package for imaging reconstruction with the robust parameter equal to zero.  
For a better sensitivity, visibility data were smoothed from an instrumental spectral resolution of $0.1 \; \mathrm{km \, s^{-1}}$ to $0.2 \; \mathrm{km \, s^{-1}}$ before making images.  
The TP image cubes were used as default images when performing the maximum entropy deconvolution.
To avoid undesired distortion in following statistical analyses, the spectral line image in each mosaic field was restored with a circular beam size equal to the solid angle of the Gaussian synthesized beam.   
The beam size of the final image is $3\farcs48$ for Field-N and $3\farcs15$ for Field-S, and the pixel size is $0\farcs3$.  
The rms noise level per channel is $27 \; \mathrm{mJy \, beam^{-1}}$ ($0.32 \; \mathrm{K}$) in Field-N and $23 \; \mathrm{mJy \, beam^{-1}}$ ($0.32 \; \mathrm{K}$) in Field-S.  
Integrated intensity maps are also generated with visibilities averaged within a velocity range of $8.0-28.0 \; \mathrm{km \, s^{-1}}$ and $8.8-28.8 \; \mathrm{km \, s^{-1}}$ for Field-N and Field-S, respectively.  
The respective rms noise level is $1.30 \; \mathrm{K \, km \, s^{-1}}$ and $1.24 \; \mathrm{K \, km \, s^{-1}}$ in Field-N and Field-S.   

\section{Results} \label{sec:results} %
\begin{figure}[h!] %
\epsscale{1.2} %
\plotone{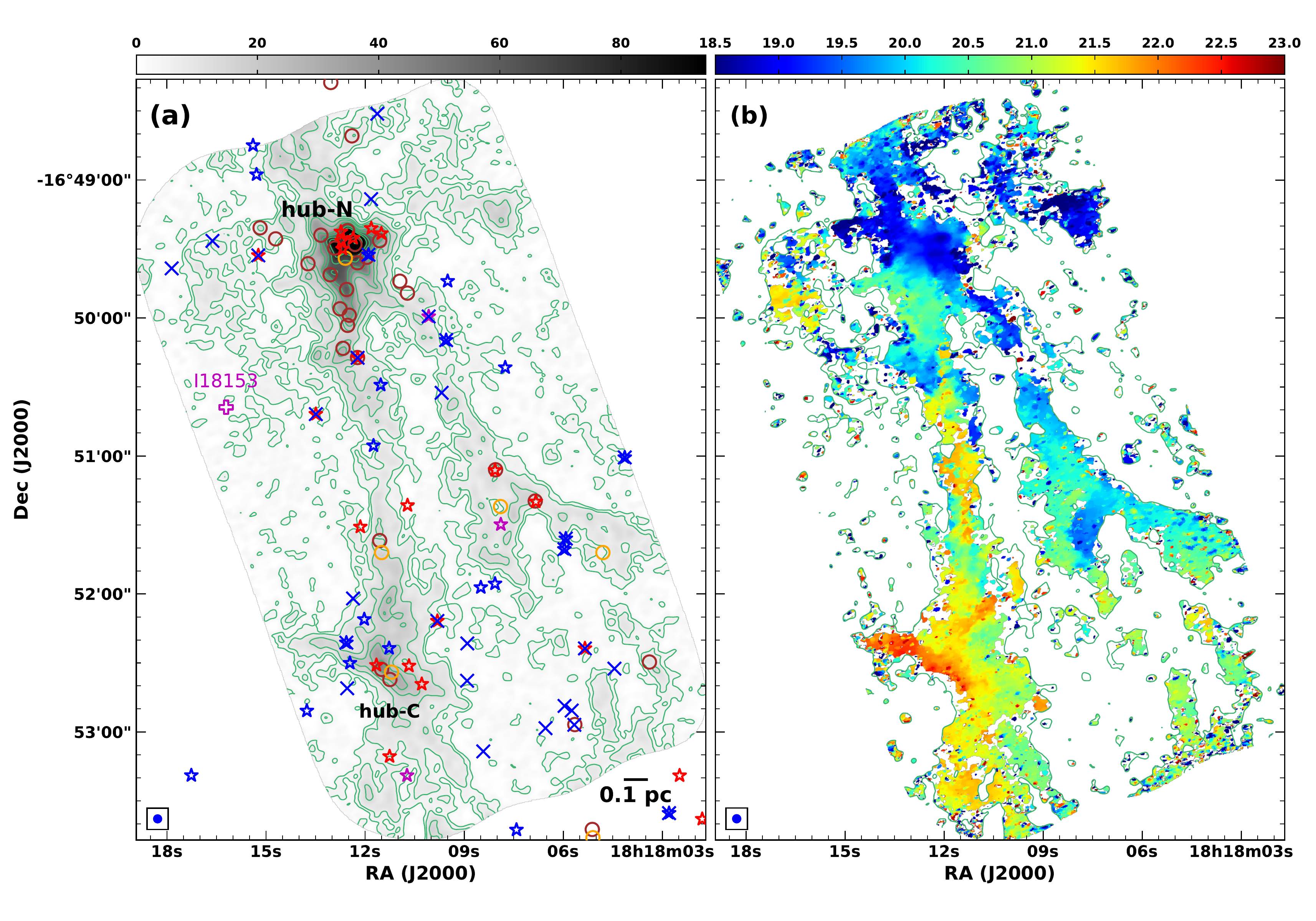} %
\caption{(a) ALMA $\mathrm{N_2H^+} \; (1-0)$ integrated intensity map of Field-N showing a remarkable filamentary morphology in dense gas.  The dominant star-forming core, hub-N, is associated with dense cores, embedded YSOs, and several prominent filaments in this region.  One more hub candidate, hub-C, shows stronger $\mathrm{N_2H^+}$ emission and is also associated with a group of YSOs and two dense cores.  Contour levels are $(3,5,10,15,20,30,40,50,60,80,100) \times \sigma$, where the rms noise level is $\sigma = 1.30 \; \mathrm{K \, km \, s^{-1}}$.   Dense cores identified in the $3 \, \mathrm{mm}$ continuum emission and dense clumps identified in the $870 \; \mu\mathrm{m}$ continuum emission  \citep{Ohashi:2016iz} are shown as brown and orange open circles, respectively.  The magenta cross marks the position of IRAS 18153$-$1651.  YSOs with $A_V > 20 \; \mathrm{mag}$ are also shown \citep{Povich:2016jm}: SED classification in Stage 0/I sources (red stars), Stage II/III (blue stars), and ambiguous (magenta stars).  Blue crosses mark the positions of X-ray sources associated with the cluster \citep{Povich:2016jm}.  
(b) Intensity-weighted velocity (moment 1) map of the isolated hyperfine component ($F_1F = 01 \rightarrow 12$) of $\mathrm{N_2H^+} \; (1-0)$ emission.  One can see a generally increasing trend in velocity from north to south.  Each filament appears in a slightly different velocity range.   
\label{fig:g14n}  }%
\end{figure}

\begin{figure}[h!] %
\epsscale{1.2} %
\plotone{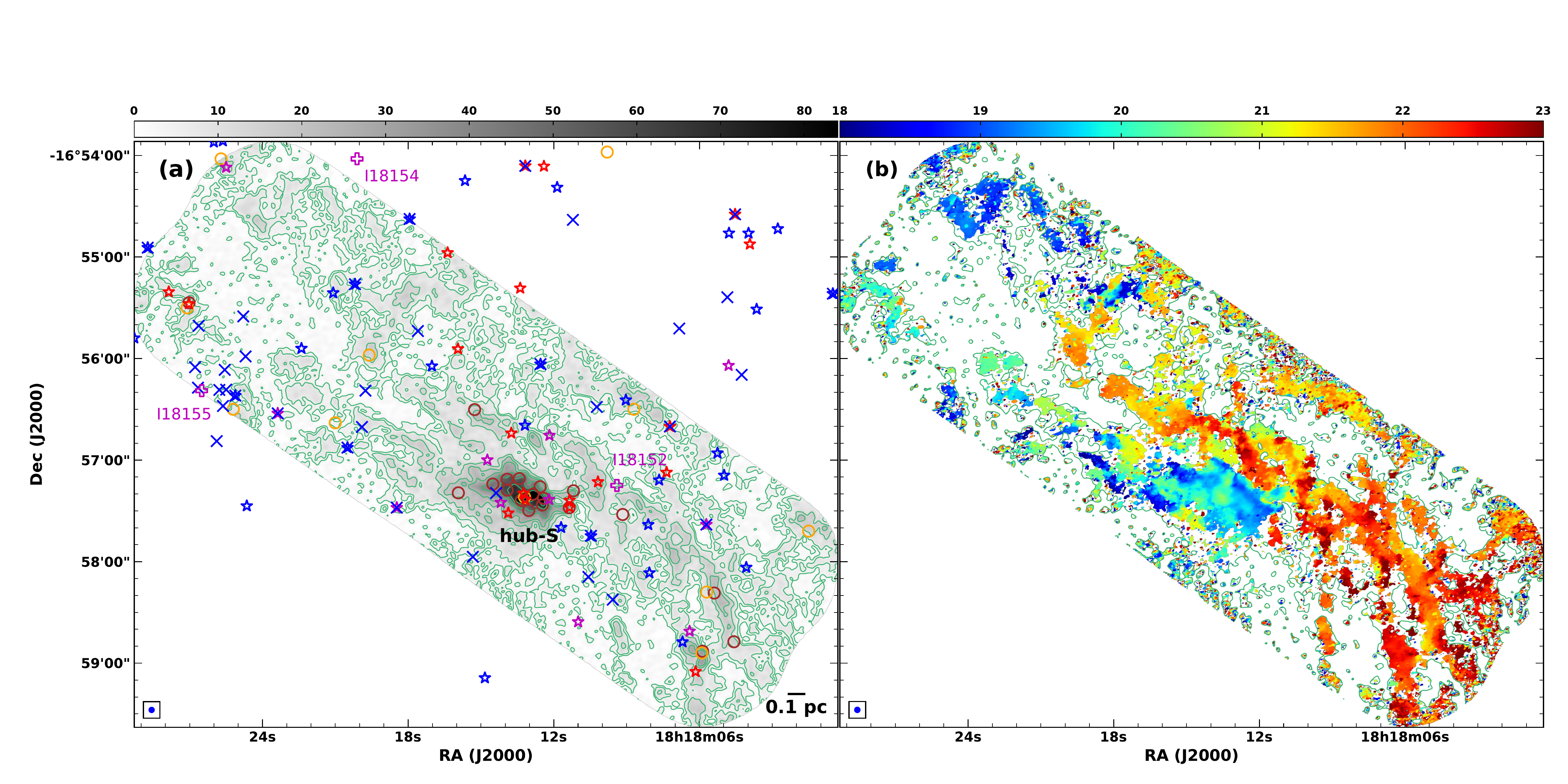} %
\caption{(a) ALMA $\mathrm{N_2H^+} \; (1-0)$ integrated intensity map of Field-S.  The dominant star-forming core, hub-S, is associated with dense cores, embedded YSOs, and several prominent filaments in this region.  Contour levels are $(3,5,10,15,20,30,40,50,60,80) \times \sigma$, where the rms noise level is $\sigma = 1.24 \; \mathrm{K \, km \, s^{-1}}$. The magenta crosses mark the positions of IRAS~18155$-$1657, IRAS~18152$-$1658, and IRAS~18154$-$1655.  All other symbols are the same as in  Fig.~\ref{fig:g14n}. 
(b) Intensity-weighted velocity (moment 1) map of the isolated hyperfine component ($F_1F = 01 \rightarrow 12$) of $\mathrm{N_2H^+} \; (1-0)$ emission.  One can see two fairly distinct velocities between the north-east part and south-west part of the cloud.  
\label{fig:g14s} } %
\end{figure}

Figures~\ref{fig:g14n} and \ref{fig:g14s} show the integrated intensity maps of the $\mathrm{N_2H^+} \; (1-0)$ emission in the two mosaic fields along with their intensity weighted velocity (moment-1) maps generated solely with the isolated hyperfine component ($F_1F = 01 \rightarrow 12$).   
The intensity weighted velocity maps are computed with a clip value of $2.5 \sigma$ over a velocity range of $17.1 - 23.1 \; \mathrm{km \, s^{-1}}$ for Field-N and $17.7 - 23.7 \; \mathrm{km \, s^{-1}}$ for Field-S. 
Similar to the ammonia emission reported by \citet{Busquet:2013ko}, the $\mathrm{N_2H^+}$ emission also shows a network of filaments, where hubs are located at the intersection of multiple filaments.    
With much improved angular resolution of $\sim 3\farcs5$ (equivalent to $0.034 \; \mathrm{pc}$), we are able to resolve structures down to their thermal Jeans length of $0.06 \; \mathrm{pc}$ (assuming a density of $10^5 \; \mathrm{cm^{-3}}$ at $10 \; \mathrm{K}$).    
Both hubs show extended, elongated structures connecting to their surrounding filaments.  
The velocity distribution in both fields show a general flow pattern (Fig.~\ref{fig:g14n}b and \ref{fig:g14s}b).  
An overall velocity gradient is clearly revealed in each mosaic field, suggesting inflow motions along filaments, most likely towards the center of gravity, where the hubs are located.  
Since our spectra show multiple velocity components in many positions, one should regard the intensity weighted velocity maps as weighted mean velocity distribution of the actual complicated kinematics in the regions.  
In Field-N, gas velocity decreases from $\sim 22.5 \; \mathrm{km \, s^{-1}}$ in the south to $\sim 19.0 \; \mathrm{km \, s^{-1}}$ in the north.  
Two fairly distinct velocity distribution are found in Field-S, where velocity increases from $\sim 19 \; \mathrm{km \, s^{-1}}$ in the north-east to $\sim 23 \; \mathrm{km \, s^{-1}}$ in the south-west.  

In general, deeply embedded YSOs \citep[red stars as stage 0/I sources;][]{Povich:2016jm} are found to be associated with hubs and filaments while evolved YSOs \citep[blue stars as stage II/III and blue crosses as X-ray sources;][]{Povich:2016jm} appear more distributed in the regions.  
Dense cores identified in the continuum studies \citep[brown open circles;][]{Ohashi:2016iz} are preferentially located in hubs, and just a few in filaments.   
This perceptible association of dense cores and deeply embedded YSOs with filaments, particular in the vicinity of the hubs, assures that filaments are part of star formation processes instead of occasional over-dense features in clouds.  
Filaments may participate in star formation in two ways: fragmentation into cores with nearly equal spacings \citep[e.g.][]{NaranjoRomero:2012uo,Wang:2011jl,Zhang:2009hw,Zhang:2015ua} or longitudinal accretion flow along the axis \citep[e.g.][]{Peretto:2013kt,Kirk:2013gq,Contreras:2016iu,Lu:2018bn}.  
Based on the $\mathrm{N_2H^+}$ gas flow motion towards the hubs and positions of the continuum dense cores not being regularly spaced (Fig.~\ref{fig:g14n} and \ref{fig:g14s}), the filaments in IRDC~G14.2 are more inclined to mass accretion into the hubs rather than fragmentation into cores.  
Yet the two aspects are not mutually exclusive and may occur simultaneously.   

In addition, a small group of YSOs and two dense cores appear to be associated with a hub candidate, hub-C, at convergence of two elongated structures with different orientations (Fig.~\ref{fig:g14n}).  
Similar to hub-N and hub-S, this candidate hub also shows warm and compact $\mathrm{NH_3} \; (2,2)$ emission \citep{Busquet:2013ko}, whose upper level energy is $65 \; \mathrm{K}$. 
The rotational temperature in hub-C ranges from $15$ to $25 \; \mathrm{K}$ (Busquet, private communication).  

\begin{figure}[h!] %
\epsscale{1.0} %
\plotone{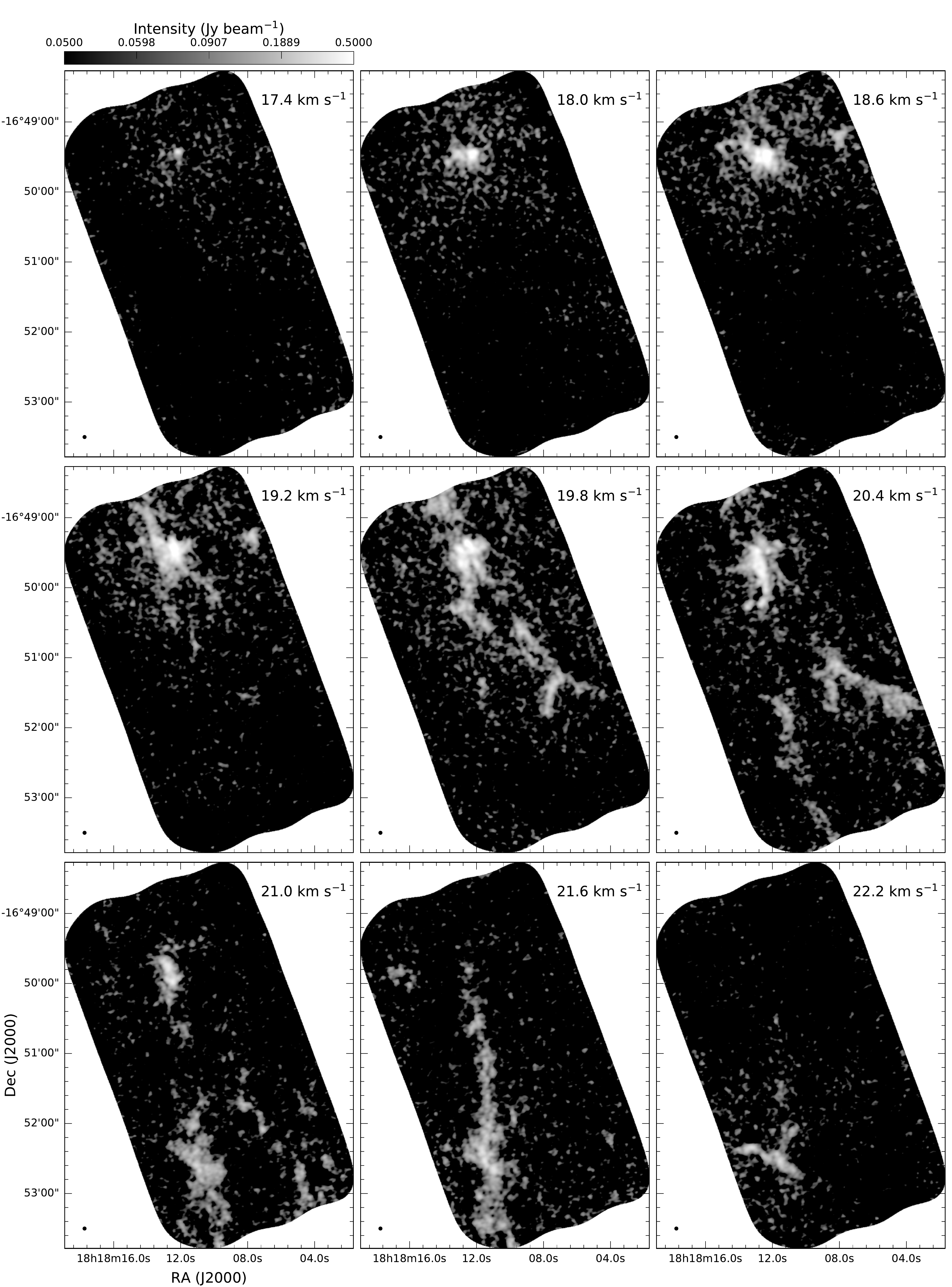} %
\caption{Velocity channel maps the isolated hyperfine component ($F_1F = 01 \rightarrow 12$) of $\mathrm{N_2H^+} \; (1-0)$ emission in Field-N with step of $0.6 \; \mathrm{km \, s^{-1}}$.  
To show filamentary structures, the grayscale is saturated in hub-N.
The two dominant filaments are well separated in space and velocity.  
The velocity channel maps used for our spectral analysis with step of $0.2 \; \mathrm{km \, s^{-1}}$ is available as a figure set (Fig.~\ref{fig:g14n_fullchan}) in the online journal.  
An animated version of the velocity channel maps in step of $0.2 \; \mathrm{km \, s^{-1}}$ is available in the online journal.  The video duration is 8~s. 
\label{fig:g14n_chan}} %
\end{figure}

\begin{figure}[h!] %
\epsscale{1.2} %
\plotone{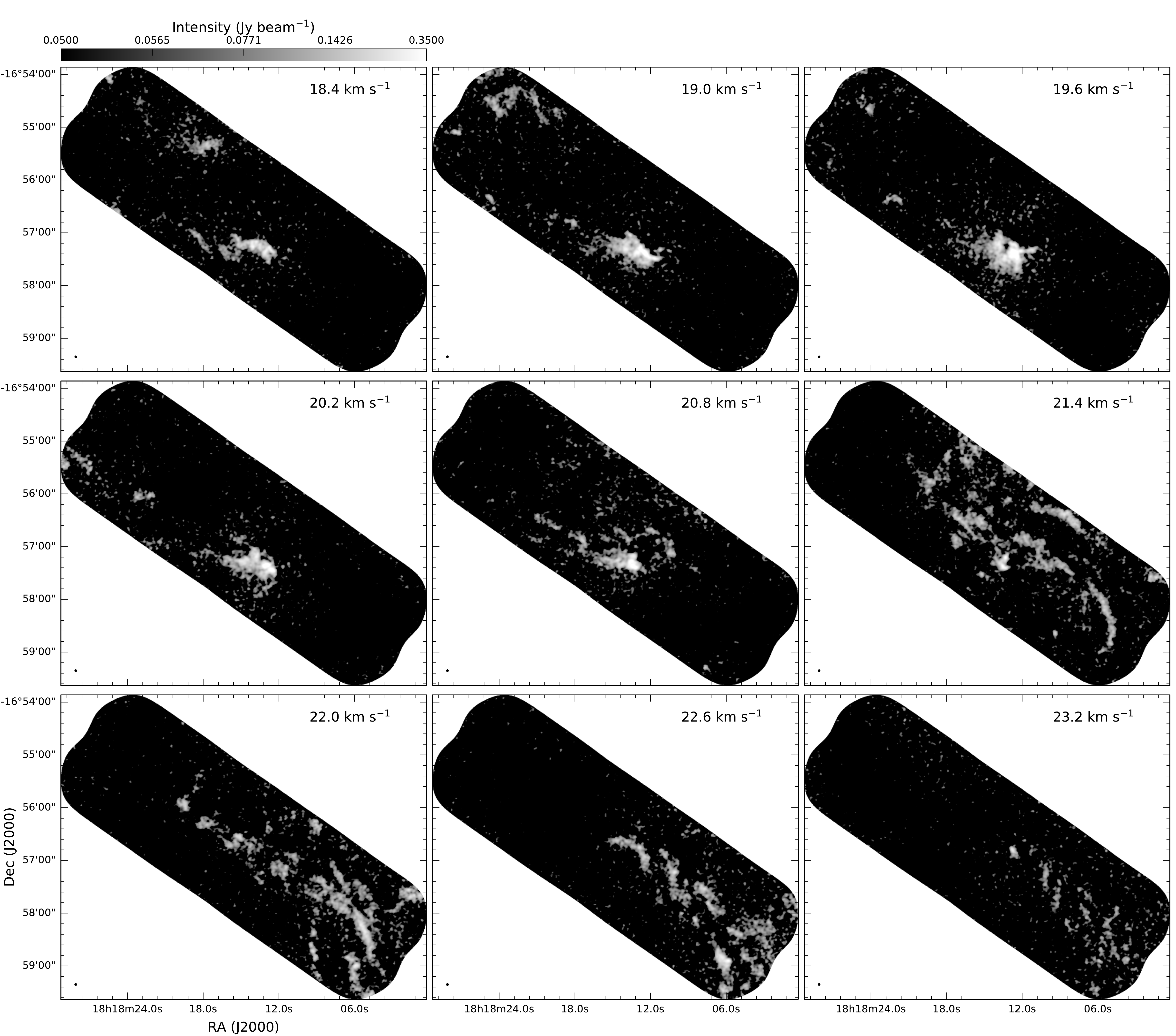} %
\caption{Velocity channel maps of the isolated hyperfine component ($F_1F = 01 \rightarrow 12$) of $\mathrm{N_2H^+} \; (1-0)$ emission in Field-S with step of $0.6 \; \mathrm{km \, s^{-1}}$.  
To show filamentary structures, the grayscale is saturated in hub-S.
Multiple filaments are likely present in this field.    
The velocity channel maps used for our spectral analysis with step of $0.2 \; \mathrm{km \, s^{-1}}$ is available as a figure set (Fig.~\ref{fig:g14s_fullchan}) in the online journal.  
An animated version of the velocity channel maps in step of $0.2 \; \mathrm{km \, s^{-1}}$ is available in the online journal.  The video duration is 8~s.  
\label{fig:g14s_chan}} %
\end{figure}

The general gas kinematics are shown in velocity channel maps with step of $0.6 \; \mathrm{km \, s^{-1}}$ (Figs.~\ref{fig:g14n_chan} and \ref{fig:g14s_chan}).  
In addition, velocity channel maps of the isolated component with a spectral resolution of  $0.2 \; \mathrm{km \, s^{-1}}$ (Figs.~\ref{fig:g14n_fullchan} and \ref{fig:g14s_fullchan}) and the corresponding movies (Figs.~\ref{fig:g14n_chan} and \ref{fig:g14s_chan}) are available in the online journal. 
A few filaments are easily identified as persistent structures across consecutive velocity channels with clear velocity gradients.  
The two prominent hubs, hub-N and hub-S, both exhibit large velocity spreads of more than $2 \; \mathrm{km \, s^{-1}}$.       
Hence we specify the spatial extent of the hubs to be regions with emission in more than 10 consecutive channels, equivalent to $2 \; \mathrm{km \, s^{-1}}$, and with intensities greater than $3\sigma$ in the isolated hyperfine component.   

\section{Analyses} \label{sec:analyses} %
\subsection{Filament Identification} \label{subsec:flmid} %
We use the publicly available filament finding package {\sf FilFinder}  \citep{Koch:2015dc}\footnote{FilFinder available online at \url{ https://github.com/e-koch/FilFinder}} skeletons.    
{\sf FilFinder} isolates filamentary structures by creating a mask with adaptive threshold, where a valid pixel must have intensity greater than the median of the neighborhood around it.  
An object also need to have an aspect ratio larger than 5 to be considered as a branch or filament.  
Each filament mask is then reduced to a skeleton using medial axis transform.    
The algorithm then drives the shortest path between each pair of end points and finds the longest path in a connected path network to be the final filament spine.    
The remaining branches are pruned leaving the dominant spine for further analyses.  
The sum of branch lengths for this dominant spine is defined to be the length of the filament, $\ell$.  

\begin{figure}[h!] %
\epsscale{1.2} %
\plotone{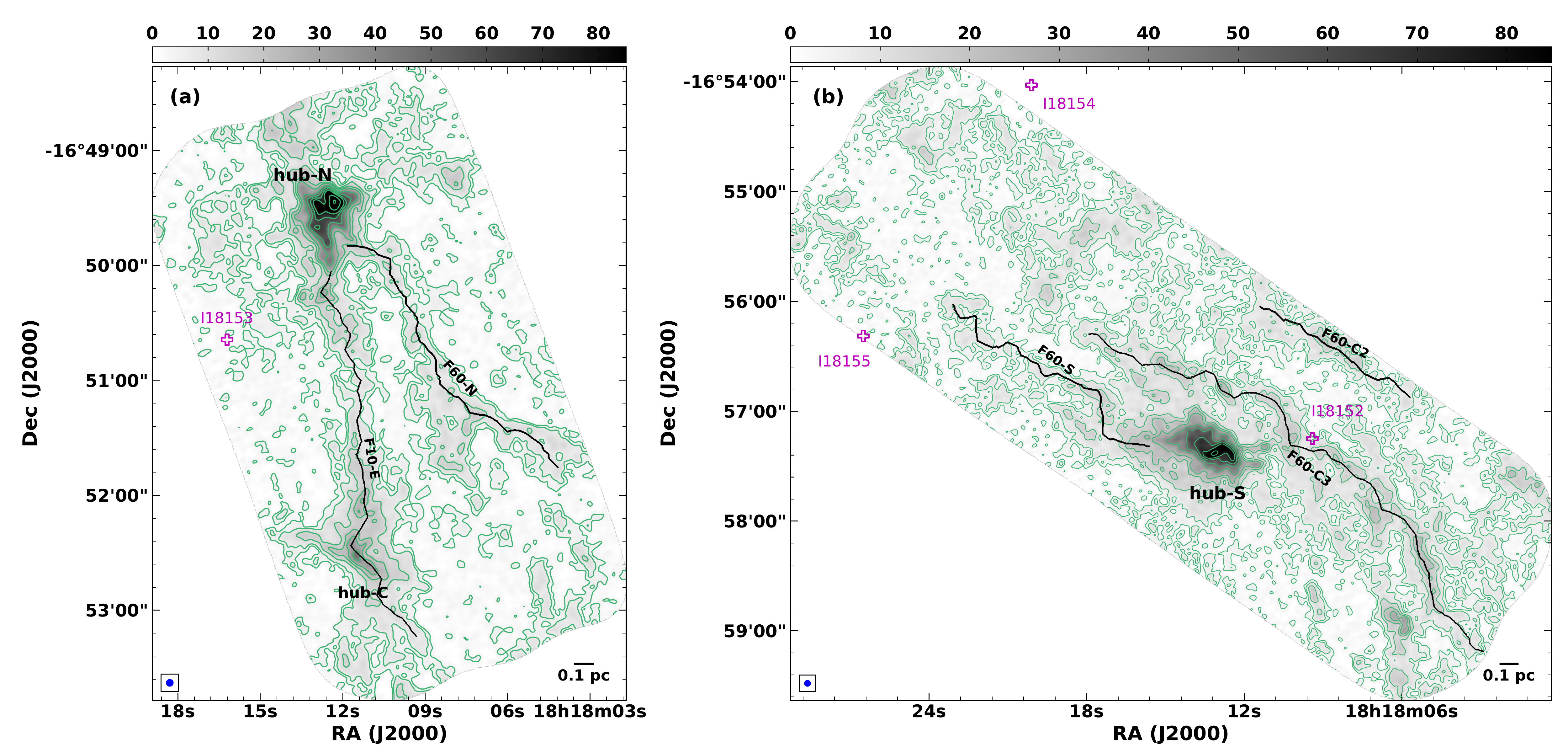} %
\caption{ (a) Filaments identified with the {\sf FilFinder} package in Field-N overlaid on the integrated intensity map.  The magenta cross marks the positions of IRAS 18153$-$1651.  Contour levels follow  those in Fig.~\ref{fig:g14n}.  (b) Identified $\mathrm{N_2H^+} \; (1-0)$ filaments in Field-S overlaid on the integrated intensity map.  The magenta crosses mark the positions of IRAS~18155$-$1657, IRAS~18152$-$1658, and IRAS~18154$-$1655.  Contour levels follow those in Fig.~\ref{fig:g14s}.  \label{fig:flm}
\label{fig:flmid}} %
\end{figure}

To extract skeletons in our mosaic fields, we first select persistent structures in consecutive velocity channels (at least 3 channels for local structures and 8 channels for the entire filament) and generate integrated intensity maps that are most optimal to individual structures.  
However, this approach restricts us to use solely the isolated hyperfine component, which does not suffer from line blending but is merely $\sim 1/9$ in the total intensity.  
Although the signal-to-noise ratio is lower, we are able to extract skeletons, which are relatively brighter features in filaments. 
The mask is prepared with a flatten percentage of 90\% and a minimum intensity to be included ({\sf glob\_thresh}) at 60\% for field-N and 40\% for field-S.  
We prune branches shorter than $1 \; \mathrm{pc}$ to avoid confusion in the margins of the mosaic fields.  
The performance of {\sf FilFinder} is fairly robust.  The spine identification does not vary drastically unless the parameters are deviated far from the default values.  
In addition, we terminate a spine when it enters one of the two prominent hubs,  
whose spatial extent is defined as a region with emission higher than $3 \sigma$ in more than $10$ channels (see Section~\ref{sec:results}).    
This is to avoid a hub connecting all its associated filaments as one whole kinetic ensemble structure.  
In total, {\sf FilFinder} identifies five $\mathrm{N_2H^+}$ filaments (Fig.~\ref{fig:flmid}): two in Field-N and three in Field-S.   
Since these filaments spatially overlap with the ammonia filaments \citep{Busquet:2013ko}, we simply follow the nomenclature to label the $\mathrm{N_2H^+}$ filaments and list their basic properties in Table~\ref{tab:flmid}.  
The length of the $\mathrm{N_2H^+}$ filaments, $\ell$, is in the range of $1.02$ to $3.22 \; \mathrm{pc}$ with a mean value of $2.0 \; \mathrm{pc}$.
We consider these projected filament lengths to be the lower limits of the actual filament lengths in 3D space. 
Note that some previously identified $\mathrm{NH_3}$ filaments are not labelled in Fig.~\ref{fig:flmid} if they are just partially present in the periphery of our mosaic fields.  
The emission contrast of a filament to its surroundings varies among filaments.   
Filament F10-E shows the largest contrast while F60-S and F60-C2 are clumpy and diffuse.  

\subsection{Filament Width} \label{subsec:flmwidth} %
Following the analyses of previous studies of filaments \citep[e.g.][]{Arzoumanian:2011ho,Palmeirim:2013da}, we analyze the integrated intensity profile of the $\mathrm{N_2H^+ \; (1-0)}$ emission with an idealized cylindrical model described by a Plummer-like function: 
\begin{equation} %
  \Sigma_p(r) = A_p \frac{\rho_c \, R_\mathrm{flat}}{\left[ 1 + (r/R_\mathrm{flat})^2 \right]^{(p-1)/2}},  
  \label{eq:plummer} %
\end{equation} %
where $\rho_c$ is the central density of the filament, $p$ the power-law exponent at large radii, $A_p$ a finite constant factor, and $R_\mathrm{flat}$ the inner flat portion of the density profile.  
The filament width is given by $w = 2 R_\mathrm{flat}$.  
In the special case of an isothermal filament in hydrostatic equilibrium, \citet{Ostriker:1964gt} found $p=4$, $A_p = \pi/2$, and $R_\mathrm{flat}$ equal to the thermal Jeans length at the center of the filament.    
Early studies with {\it Herschel} continuum data found $p=2$ and $w = 0.1 \; \mathrm{pc}$ \citep{Arzoumanian:2013gp,Palmeirim:2013da}.   
This $r^{-2}$ density profile has also been reproduced in numerical models of filamentary clouds with helical magnetic fields and turbulent pressure of the surrounding interstellar medium \citep{Fiege:2000ds}.  
Yet steeper density profile with $p = 2.7 - 5.1$ have also been reported in a number of filaments  \citep{Nutter:2008hf,Hacar:2011ij,Pineda:2011ds,Monsch:2018gb}.  

To find the filament width, $w$, we apply the publicly available filament profile builder package {\sf RadFil} \citep{Zucker:2018fp,Zucker:2018eg}\footnote{RadFil available online at \url{https://github.com/catherinezucker/radfil}} to the integrated intensity maps of the $\mathrm{N_2H^+} \; (1-0)$ emission. 
Given a filament spine together with a filament mask, {\sf RadFil} first smooth the input spine pixels to a continuous version of the spine, i.e. a basis spline, and then create cuts based on the positions and the first derivative of the basis spine to build intensity profile across the filament.    
The algorithm allows shifting the profile by searching for the pixel with the peak intensity along each cut but inside the filament mask.  
Once the radial profiles along all the cuts are computed, {\sf RadFil} fits a given profile function, such as the Plummer function, on the average profile of the entire ensemble of cuts.  
We produce a mask of $0.1 \; \mathrm{pc}$ width following each individual filament spine to confine the search for the peak intensity pixels so the algorithm will not be confused by nearby branches or filaments.   
Cuts are taken every $12$ pixels ({\sf samp\_int=12}), which are roughly equal to the beam size.  
{\sf RadFil} also allows a background emission subtraction before performing the profile fit.  
A first-order polynomial fit is applied to the background emission model for proper subtraction. 
The background model is obtained with data in regions between $0.2 - 0.8 \; \mathrm{pc}$ from the spine.  The range is selected to obtain most of the available data on the two sides of individual filaments.  
We also mask out hubs to reduce the bias in the background subtraction when computing the profiles along cuts.  

\begin{figure}[h!] %
\epsscale{0.9} %
\plotone{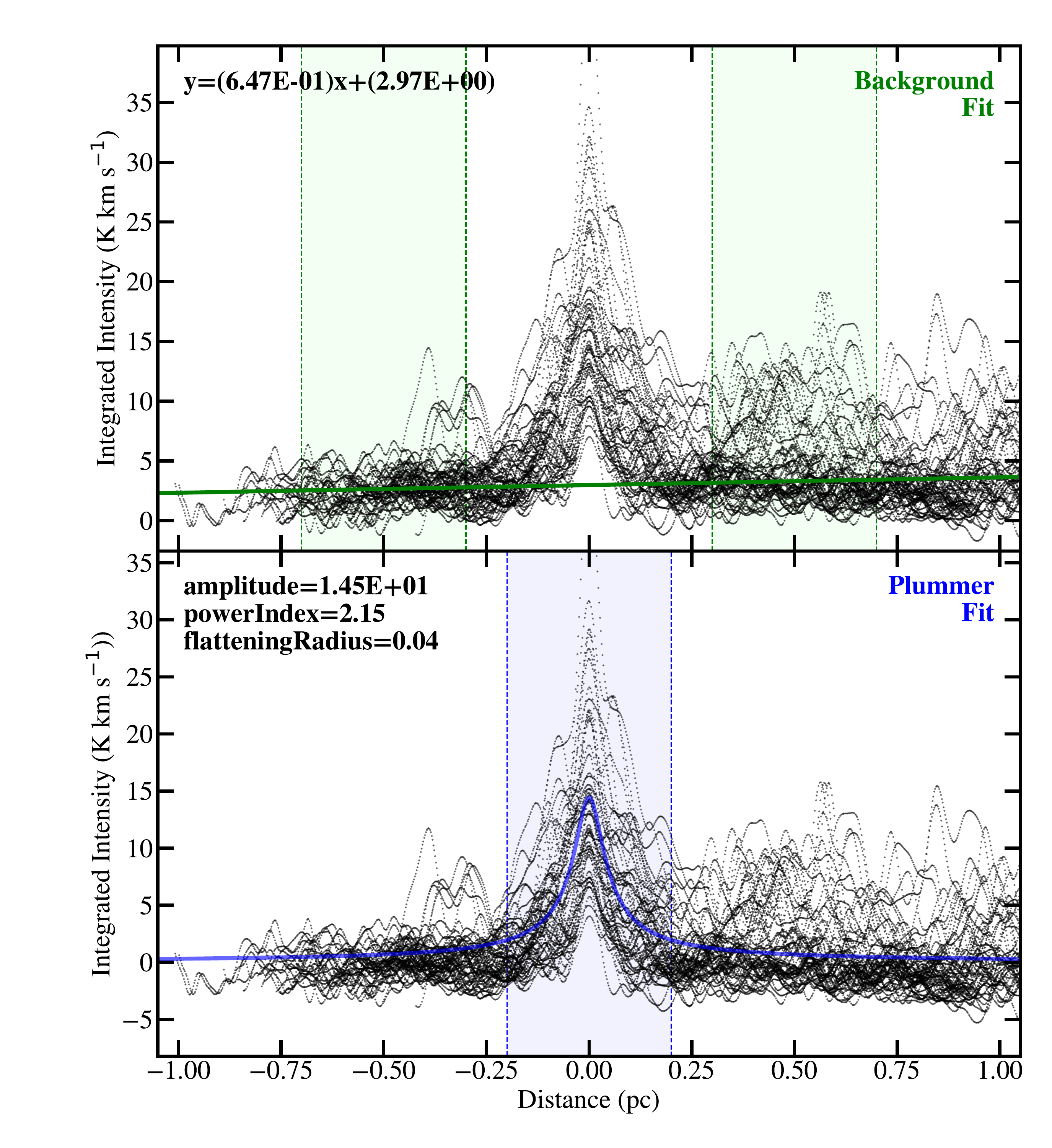} %
\caption{
Plummer function fit with the integrated intensity cuts across filament F10-E using the Python package {\sf RadFil}.  (Top) Background model with a first-order polynomial using data between $0.3$ and $0.7 \; \mathrm{pc}$ from the spine (green shaded regions bounded by vertical green dash line).   Thick green line shows the best-fit background.  Thin black lines show the profiles of the entire ensemble of cuts. (Bottom) The best-fit Plummer profile (thick blue line) to the background-subtracted profiles (thin black lines) within $0.2 \; \mathrm{pc}$ from the spin.    
\label{fig:radfil}} %
\end{figure}

The widths from the optimized fits are given in Table~\ref{tab:flmid}.  
As an example, the profile fit along the spine of filament F10-E is shown in Fig.~\ref{fig:radfil}.  
The filament widths are in the range of $0.05$ to $0.09 \; \mathrm{pc}$ with a mean value of $0.07 \; \mathrm{pc}$, which is smaller but comparable to the characteristic width of $0.1 \; \mathrm{pc}$ reported in the previous {\it Herschel} studies with dust emission \citep[e.g.][]{Arzoumanian:2011ho,Palmeirim:2013da,Arzoumanian:2019ky}.
Narrow widths down to $\sim 0.02 \; \mathrm{pc}$ have also been reported in a number of filaments observed with molecular gas \citep[e.g.][]{Pineda:2011ds,Hacar:2018eh,Monsch:2018gb}.   
Meanwhile, theoretical studies have pointed out possible bias introduced by the method for interpreting the data.  
Filaments are made up of pre-existing short sub-filaments, and the widths may have a broader distribution instead of being a constant \citep{Smith:2014tx}.
The measured width is also likely to be affected by the choice of parameters \citep{Panopoulou:2017kg}.  

\begin{deluxetable}{lCCCC}[h!] %
\tablecaption{Filament Length and Width \label{tab:flmid}} %
\tablecolumns{5} %
\tablewidth{0pt} %
\tablehead{
\colhead{Filament} & \colhead{$\ell$ (pc)\tablenotemark{a}} & \colhead{$w$ (pc)} & \colhead{$p$}  & \colhead{$\upsilon$ ($\mathrm{km \, s^{-1}}$)\tablenotemark{b}}  } %
\startdata %
\sidehead{Field-N} %
\cline{1-1} %
F10-E & 2.26 & 0.07 \pm 0.05 & 2.2 \pm 0.6 & 19.7-22.9 \\
F60-N & 1.81 & 0.09 \pm 0.07 & 3.6 \pm 2.5 & 18.9-21.2 \\
\hline 
\sidehead{Field-S} %
\cline{1-1}
F60-C3 & 3.22 & 0.05 \pm 0.03 & 2.5 \pm 0.6 & 20.9-23.3 \\
F60-S & 1.67 & 0.05 \pm 0.06 & 2.5 \pm 1.5 & 18.1-21.1 \\
F60-C2 & 1.02 & 0.07 \pm 0.08 & 2.5 \pm 1.7 & 20.9-22.5 \\
\enddata %
\tablenotetext{a}{Filament projected length on the sky regarded as the minimum length.} %
\tablenotetext{b}{Velocity range used to identify the filament.} %
\end{deluxetable} %

\subsection{Spectral Fits for Hyperfine Structures} \label{subsec:spfit} %
We derive the kinematics pixel by pixel with a sophisticated spectral model that accounts for multiple velocity components as well as optical depth and line blending.  
Along each line of sight (each pixel), our algorithm allows multiple velocity components in the model to interpret the expected complicated kinematic structure \citep{Busquet:2013ko}.  
For every velocity component, the spectrum is comprised of seven hyperfine components. 
All the velocity components are assumed to be in full thermalization for the $\mathrm{N_2H^+} \; (1-0)$ line at a single excitation temperature, $T_g$, whose value is adopted from the dust temperature, $T_d$, obtained by the iterative spectral energy distribution (SED) analysis of dust continuum emission \citep{Lin:2017bv}.  
The gas temperature, $T_g$, is expected to be well coupled to the dust temperature, $T_d$, at densities above $\sim 3 \times 10^4 \; \mathrm{cm^{-3}}$ \citep{Galli:2002ep}, applicable to regions traced by $\mathrm{N_2H^+} \; (1-0)$ (see Sect.~\ref{subsec:inflow}).    
The dust temperature map has a full spatial coverage comparing to the $\mathrm{NH_3}$ temperature map so it is used for the current study.   
Yet both temperature maps were observed with lower angular resolutions so $T_d$ is considered to be an mean temperature over a region of 10\arcsec, equivalent to $0.1 \; \mathrm{pc}$.  

\begin{figure}[h!] %
\epsscale{1.2} %
\plotone{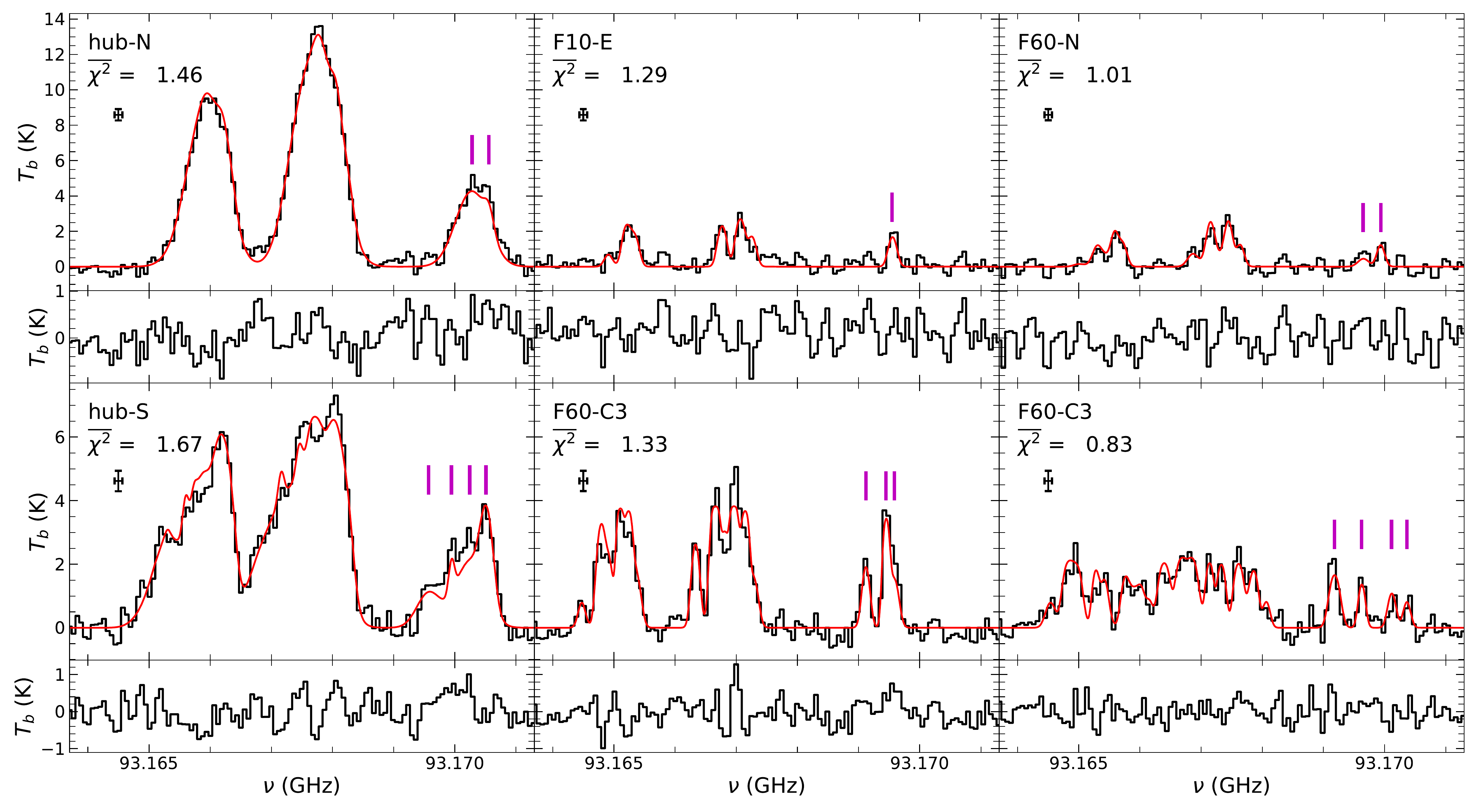} %
\caption{Observed spectrum (black histogram) with the spectral fit (red curve) and the residuals after subtracting the fit (immediately below the spectral plot) for a few selected pixels.  The magenta bars indicate the frequency of the isolated hyperfine component ($\nu_0 = 93.176252 \; \mathrm{GHz}$) shifted to the velocity components in the fit.  The associated hubs/filaments of the pixels are indicated in the upper-left corner along with the $\overline{\chi^2}$ value for the fit.  
The cross with error bars indicates the rms uncertainty ($\sigma = 0.32 \; \mathrm{K}$) in brightness temperature and the velocity channel width of $0.2 \; \mathrm{km \, s^{-1}}$.  \label{fig:spfit}} %
\end{figure}

The spectrum of each velocity component is determined by three parameters: the projected velocity, $\upsilon_i$, the line width, $\Delta \upsilon_i$, and the column density, $N_i$.  
Radiative transfer is solved to obtain the model spectrum, which is then rescaled by a constant beam filling factor, $f_b$, for all velocity components.  
Together with $f_b$, a model spectrum of $\Upsilon$ velocity components has a total of $(1+3 \Upsilon)$ free parameters for optimization.
We optimize the model spectrum by minimizing the reduced $\chi^2$ value, $\overline{\chi^2}$, which is normalized to the degrees of freedom, $\mathrm{n_\mathrm{dof}}$, and has an expectation value of~1. 
The  $\overline{\chi^2}$ is given by
\begin{equation} %
\overline{\chi^2} \equiv \frac{\chi^2}{n_\mathrm{dof}} = \frac{\chi^2}{n_\mathrm{data} - n_\mathrm{par}}, 
\end{equation} %
where $n_\mathrm{data}$ is the number of data points and $n_\mathrm{par}$ is the number of fitted parameters,  $n_\mathrm{par} = 1+3\Upsilon$.  
When selecting the final solution to present the working pixel, we exclude solutions with any velocity component of spectral peak lower than $2\sigma$. 
To avoid excessive over-modeling, we also require separation between any two velocity components to be larger than 2 channels, i.e. $0.4 \; \mathrm{km \, s^{-1}}$. 
Further rejection of velocity components with spectral peaks below $3.5\sigma$ is applied to avoid poorly constrained fits, similar to the criterion used in the literature \citep[e.g.][]{Kirk:2013gq,Hacar:2018eh}.      
Fig.~\ref{fig:spfit} shows examples of our spectral fits with the observed spectra, including pixels in the hubs and filaments. 
In total, $\sim 1/3$ of the spectra display multiple velocity components.  
Details of the fitting algorithm and procedure are described in Appendix~\ref{sec:hfcfit}.  

Further inspection reveals a few limitations of our spectral models, mostly caused by over-simplified assumptions.    
Our spectral model assumes single temperature for all velocity components along line of sight, which is not valid in regions with significant internal heating, particularly in hub-N and hub-S.  
As a result, the $\overline{\chi^2}$ values are fairly large in both hubs (Fig.~\ref{fig:rcs_maps}).   
Note that we do not use spectral fits in hub-N nor hub-S for later analysis.  
Meanwhile, we also assume a single beam filling factor to reduce $n_\mathrm{par}$.   
This is not a good approximation in transition zones where multiple velocity components are involved with varying emission fractions within one single beam.  
Because of the criterion for velocity separation between any two velocity components, the selection will favor single component with a larger line width in convergent regions of multiple velocity components.   

\subsection{Velocity Components Associated with Filaments} %
Once the kinematics are derived from spectral fitting, we then try to find velocity components associated with each individual filaments in the position-position-velocity (PPV) space.  
We apply the friends-of-friends (FoF) method, one of the widely used techniques to identify groups of galaxies in galaxy redshift surveys, initially presented by \citet{Huchra:1982ed}.
 This method has recently been used to study filamentary cloud structures in  molecular line observations \citep[e.g.][]{Hacar:2013tq,Henshaw:2014ib,Hacar:2018eh}.   
The FoF is an algorithm to establish membership in a group by identifying ``friends'' of an existent group member based on their linking lengths within predetermined thresholds.   
Such a linking length criterion results in a pairwise identification that is commutative.  
If member~1 finds member~2 a friend, member~2 also finds member~1 a friend.    
In the PPV space, the linking length is usually specified by the combination of projected separation and velocity difference.   
To start the process, one usually assigns a first member, the seed, of a group.  
The algorithm then searches for its ``friends,'' which are neighbors within predetermined thresholds for the linking length.   
After including the friends of the seed, the search continues iteratively to find and include friends of the newly identified members until no more friends can be found in the catalog.  
At this point, the members of the group are determined. 

\begin{figure}[h!] %
\epsscale{1.2} %
\plotone{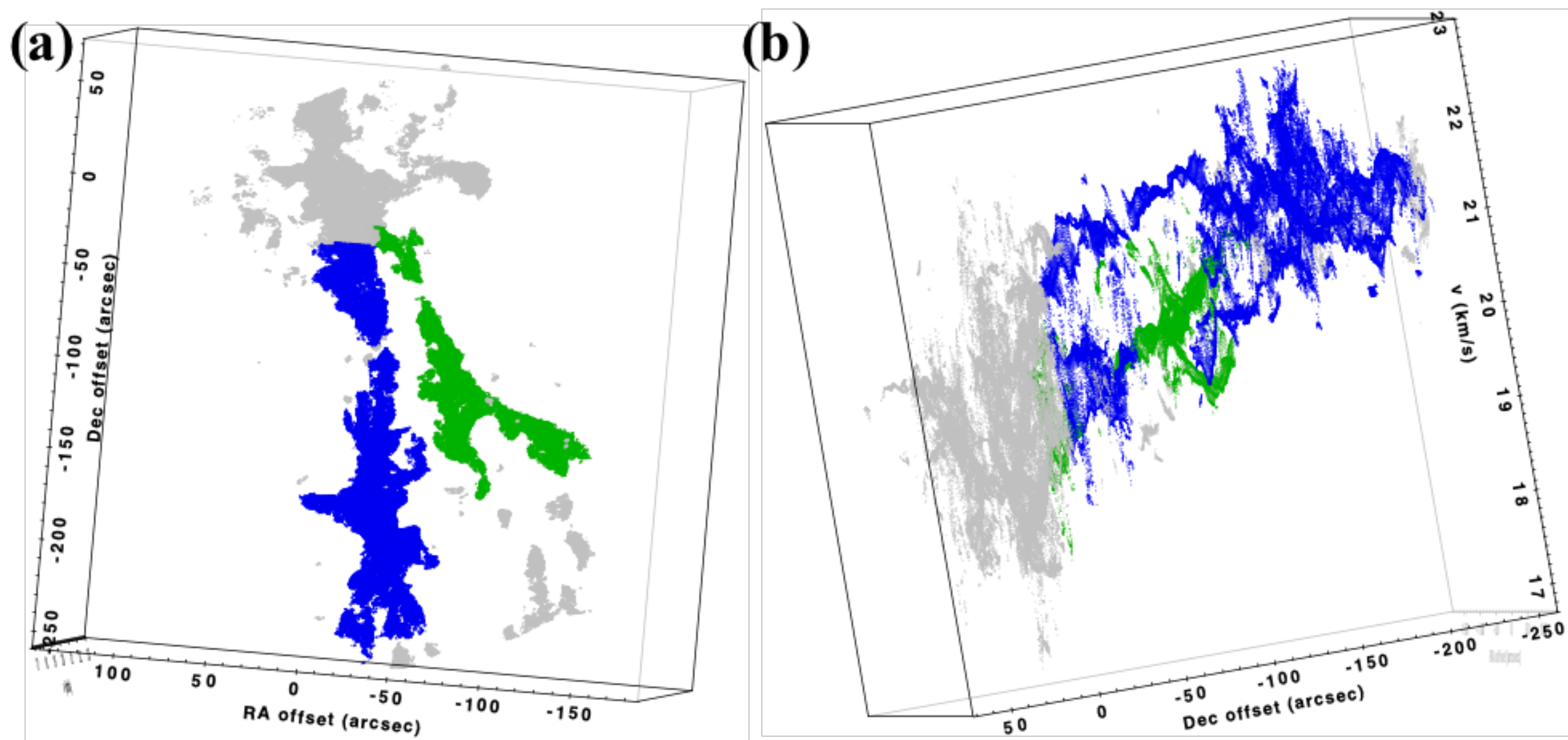} %
\caption{Selected views of position-position-velocity (PPV) distribution for the derived kinematics in Field-N.  Blue dots are components associated with filament F10-E while green dots are those associated with F60-N.  Gray dots indicate velocity components in hub-N and unclassified structures.  (a) viewing from the South.  (b) viewing from the east at an angle nearly perpendicular to the velocity axis.  A velocity gradient on parsec scale is clearly detected.    
More velocity components are present in hub-N than those in filaments.  
At $1.98 \; \mathrm{kpc}$, a structure of $1 \; \mathrm{pc}$ long subtends an angular size of $104\arcsec$. An animated version of this figure is available in the online journal as a 3D flyby.  The video duration is 15~s.  
\label{fig:g14n_ppv}} %
\end{figure}

\begin{figure}[h!] %
\epsscale{1.2} %
\plotone{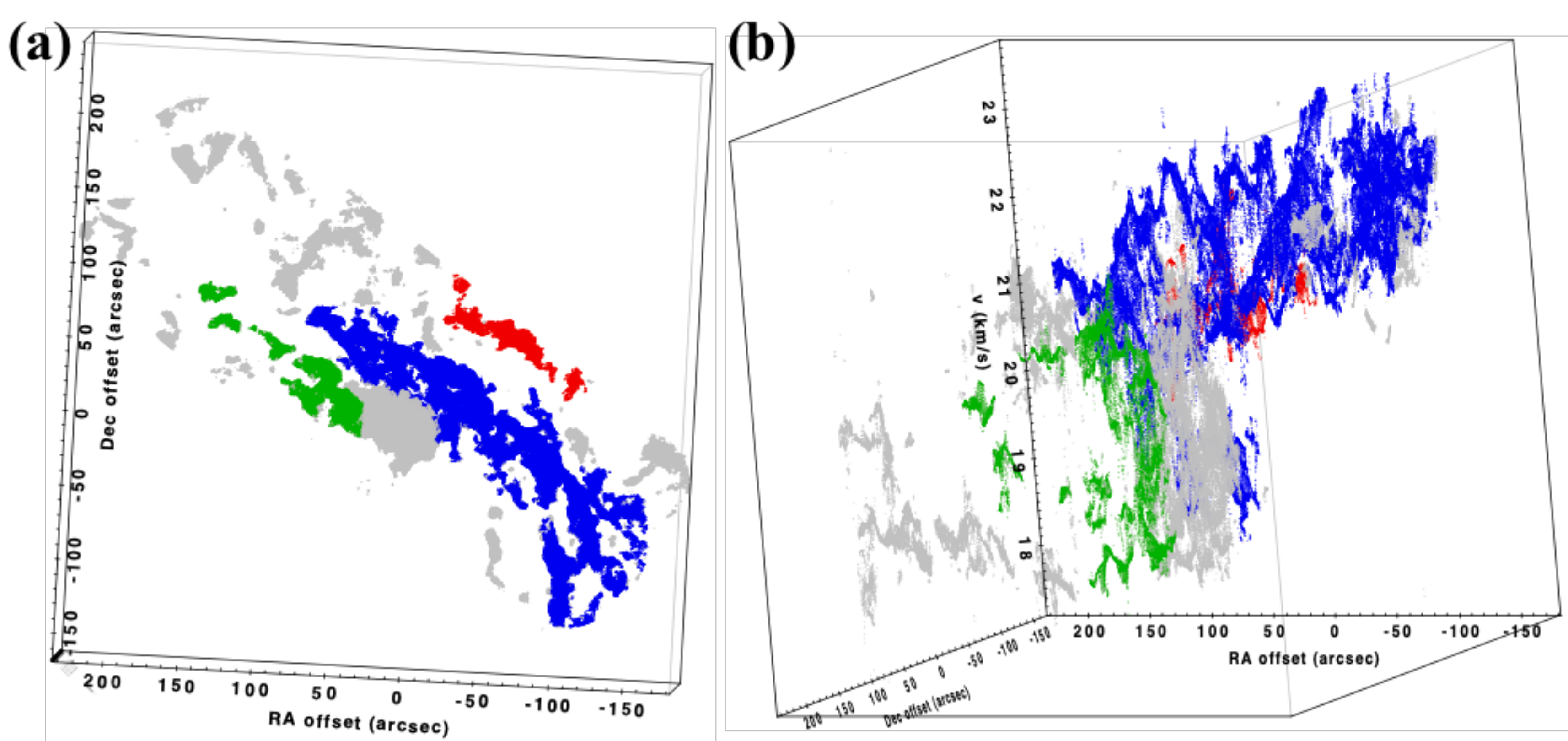}%
\caption{Selected views of position-position-velocity (PPV) distribution for the derived kinematics in Field-S.  Blue dots are components associated with filament F60-C3, green dots with F60-S, and red dots with F60-C2.  Gray dots indicate velocity components in hub-S and unclassified structures.   (a) A view from the South.  (b)  A view from the south-east at an angle nearly perpendicular to the velocity axis.  A velocity gradient on parsec scale in the filament F60-C3 is clearly detected.  More velocity components are present in hub-S than those in filaments.  At $1.98 \; \mathrm{kpc}$, a structure of $1 \; \mathrm{pc}$ long subtends an angular size of $104\arcsec$. An animated version of this figure is available in the online journal as a 3D flyby.  The video duration is 15~s. 
\label{fig:g14s_ppv}} %
\end{figure}

Given the limited sensitivity and spectral resolution, we simply search for relevant velocity components in each individual filament without further differentiating substructures in any filament.  
The thresholds are chosen to reflect the limitation of our observations.  
We consider a pair of PPV components to be friends if their angular separation is within half a beam and their velocity difference is less than two channels, i.e. $0.4 \; \mathrm{km \, s^{-1}}$. 
In addition, we set a boundary at the end point of the spine to separate filaments from the two prominent hubs; otherwise, all the filaments connected with a hub will become one single group.  
To identify PPV components associated with a filament, we assign pixels in the spine to be seeds and apply the FoF method to find members.  
Multiple seeds are assigned for a filament when regions with spectral fits that do not fully trace the spine.     
The derived PPV distribution and the components associated with each filament are shown in Fig~\ref{fig:g14n_ppv} and \ref{fig:g14s_ppv} for Field-N and Field-S, respectively.  
Animations showing the PPV distribution rotating about a fixed position are available in the online journal.  

\begin{figure}[h!] %
\epsscale{1.2} %
\plotone{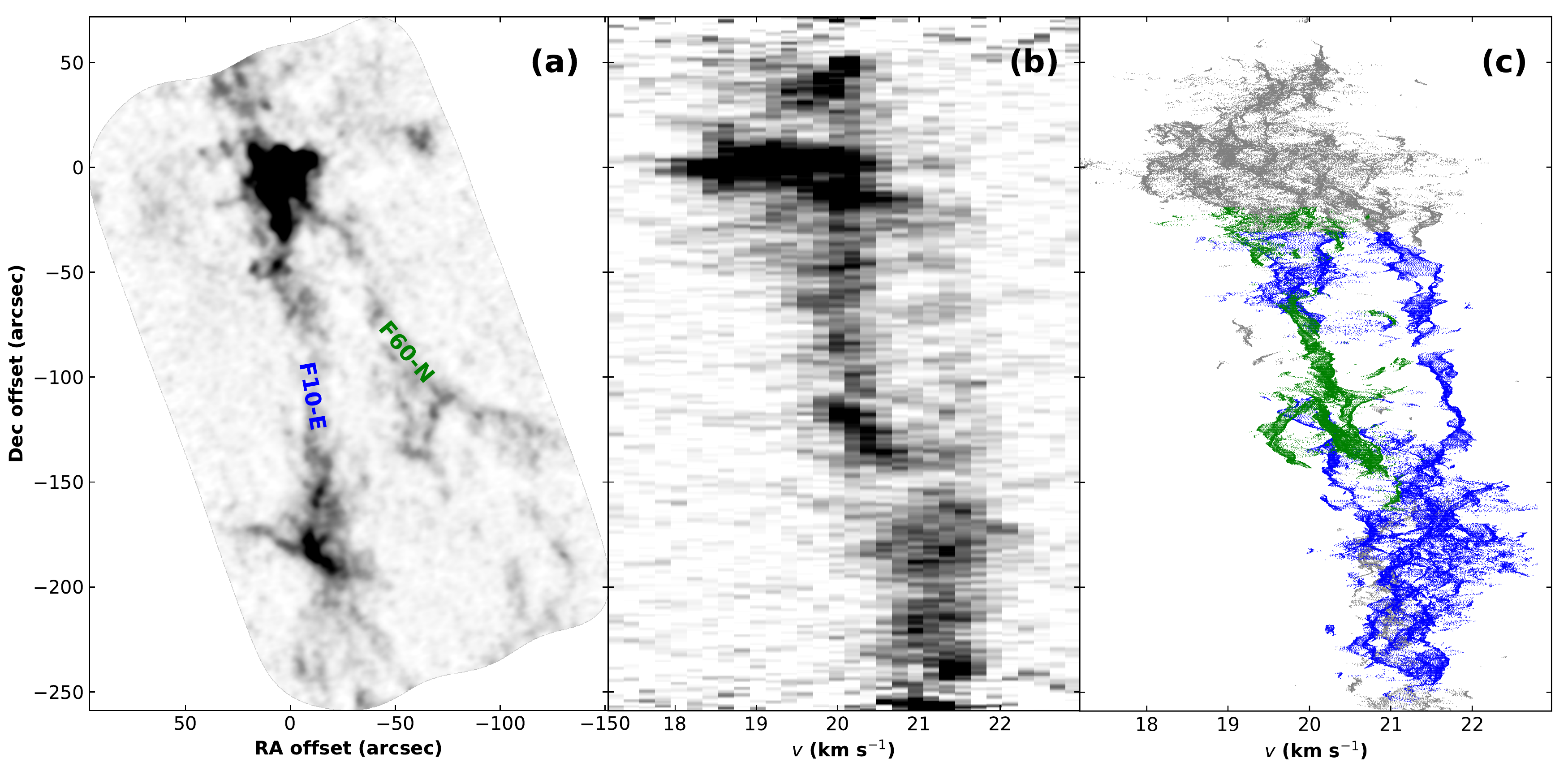} %
\caption{Gas kinematics as a function of declination in Field-N.  
(a) Integrated intensity map of Field-N with labels for identified filaments.  
(b) Position-velocity plot of averaged spectra of the isolated component of the $\mathrm{N_2H^+} \; (1-0)$ emission as a function of declination.  Spectra are averaged along the axis of right accession.  
(c) Similar position-velocity plot to (b) but for velocity determined in the hyperfine spectral fits.  
Blue points are velocity components associated with filament F10-E and green points are with filament F60-N.  
Gray dots indicate velocity components in hub-N and unclassified structures.  
\label{fig:g14n_pv}} %
\end{figure}

\begin{figure}[h!] %
\epsscale{0.65} %
\plotone{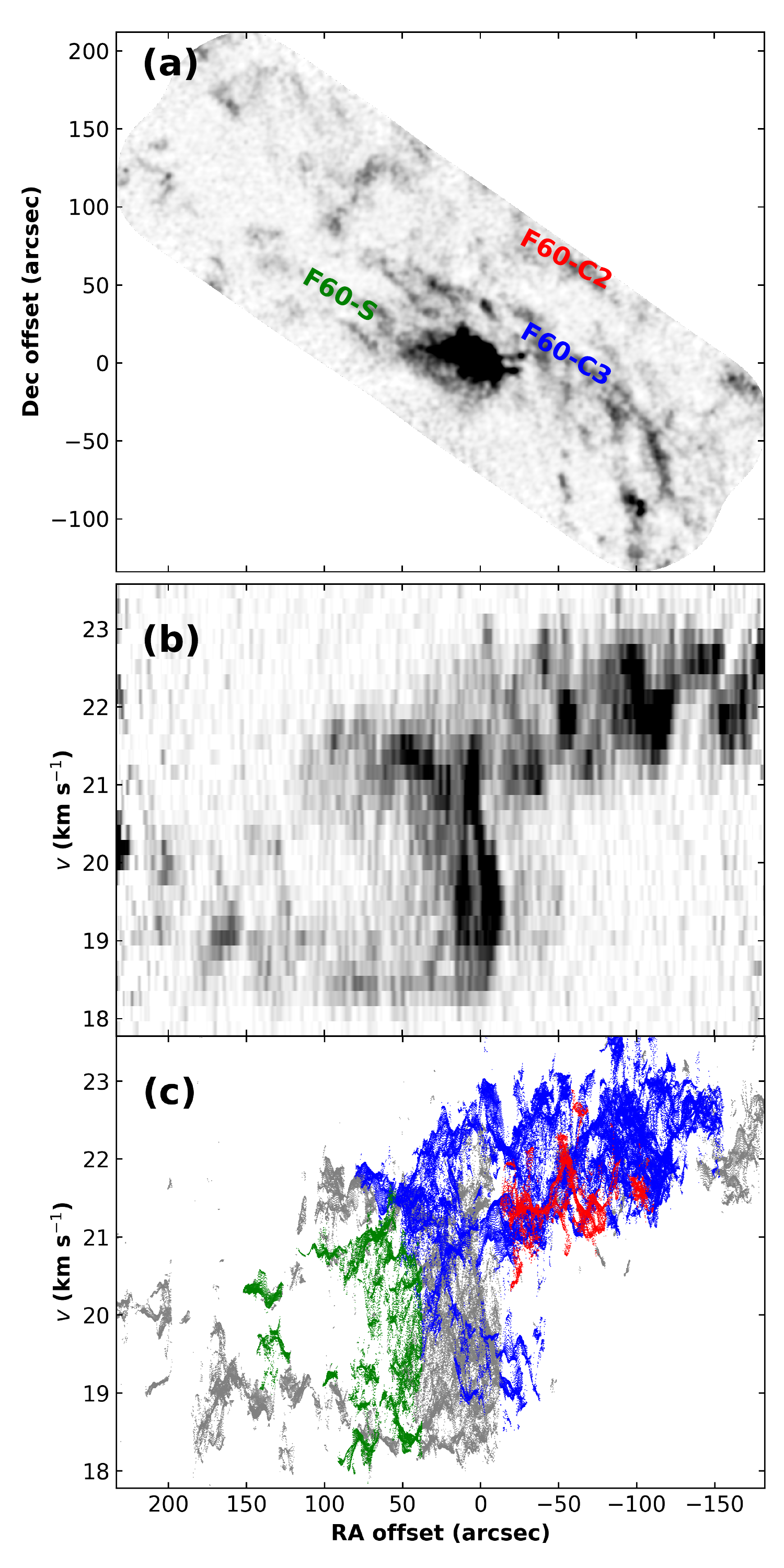} %
\caption{Gas kinematics as a function of right accession in Field-S.  
(a) Integrated intensity map of Field-S with labels for identified filaments.  
(b) Position-velocity plot of averaged spectra of the isolated component of the $\mathrm{N_2H^+} \; (1-0)$ emission as a function of right accession.  Spectra are averaged along the axis of declination.  
(c) Similar position-velocity plot to (b) but for velocity determined in the hyperfine spectral fits.  
Blue points are velocity components associated with filament F60-C3, green points with filament F60-S, and red points with F60-C2.    
Gray dots indicate velocity components in hub-S and unclassified structures. 
\label{fig:g14s_pv}} %
\end{figure}

 We also compare the derived kinematics with the averaged velocity plot of the isolated hyperfine component of the $\mathrm{N_2H^+} \; (1-0)$ emission along a given coordinate axis (Fig.~\ref{fig:g14n_pv} and \ref{fig:g14s_pv}).  
In Field-N, a general velocity gradient is present mainly along the north-south direction so we average spectra along right accession to examine the mean velocity pattern as a function of declination (Fig.~\ref{fig:g14n_pv}b).  
The velocity of all the components in the PPV space is projected as a function of declination (Fig.~\ref{fig:g14n_pv}c), where colors indicate components in different filaments.    
In Field-S, a general velocity gradient is roughly along east-west direction so spectra are averaged along the declination axis (Fig.~\ref{fig:g14s_pv}b).  
The velocity of all the PPV components is projected as a function of right accession (Fig.~\ref{fig:g14s_pv}c), where components associate with the three filaments are shown in three colors.    
In general, the derived velocity distribution agrees very well with the observed velocity pattern.  
A slow velocity gradient is present along several filaments, especially in filaments F10-E, F60-N, and F60-C3.  
Note that kinematics at the boundary between hub-S and filaments F60-C3 and F60-S are very complicated such that a perfect separation is not realistic.  

\subsection{Basic Properties of N$_2$H$^+$ Filaments} %
In general, the velocity gradient and dispersion along the filaments are better traced with the $\mathrm{N_2H^+} \; (1-0)$ line, which highlights the inner dense portion of the filaments better than the  $\mathrm{NH_3} \; (1,1)$ line does.  
This is likely due to the lower upper-level energy and higher critical density of $\mathrm{N_2H^+} \; (1-0)$. 
In addition, $\mathrm{N_2H^+}$ is known to quickly react with $\mathrm{CO}$ to form $\mathrm{HCO^+}$ in outflow regions \citep{Lee:2004fi,Jorgensen:2004kz,Busquet:2011yw,Chen:2011bo}.  
Unlike $\mathrm{NH_3}$ that may be excited by outflow shocks \citep{Zhang:1999dc}, $\mathrm{N_2H^+}$ preferentially traces dense and quiescent gas. 

Once the association with a filament is determined, we analyze the physical properties of a filament using data within $0.1 \; \mathrm{pc}$ from its spine to avoid confusion from branches.    
The width of the mask is based on the angular resolution of the column density map \citep{Lin:2016fe} that has a lower spatial resolution of $0.1 \; \mathrm{pc}$.     
Since the $\mathrm{N_2H^+} \; (1-0)$ emission may have moderate optical depth and varying abundance comparing to dust continuum emission, we estimate the mass in the filaments by linearly interpolating a column density map, $N_\mathrm{H_2}$, derived from the iterative SED analysis with a spatial resolution of $0.1 \; \mathrm{pc}$  \citep[Fig.~2 in][]{Lin:2017bv}.  
The filament total mass, $M$, is computed by integrating the column density within $0.1 \; \mathrm{pc}$ from the spine.    
One important parameter for a filament is its linear mass, which is the total mass normalized by the filament length, $M_\ell = M/\ell$.
The linear mass of our observed filaments is in the range of $75$ to $116 \; M_\sun \, \mathrm{pc^{-1}}$ with a mean value of $92 \; M_\sun \, \mathrm{pc^{-1}}$ (Table~\ref{tab:flm}).  
In general, filaments in Field-N are more massive than filaments in Field-S. 

\begin{deluxetable}{lCCCCCChCC}[h!] %
\tablecaption{Physical Properties of Filaments \label{tab:flm}} %
\tablecolumns{10} %
\tablewidth{0pt} %
\tablehead{
\colhead{Filament} & \colhead{$M_\ell$} & \colhead{$M_{\ell,\mathrm{vir}}$} & \colhead{$N_\mathrm{sub}$} & \colhead{$\alpha_\mathrm{vir}$} & \colhead{$\sigma_\mathrm{nt}/c_s$} & \colhead{$\nabla \upsilon_\mathrm{obs}$} & 
\nocolhead{$\upsilon_\mathrm{obs}$} & \colhead{$\dot{M}_\mathrm{obs}$} & \colhead{$\tau$ \tablenotemark{a}} \\
\colhead{} & \colhead{($M_\sun \, \mathrm{pc^{-1}}$)} &  \colhead{($M_\sun \, \mathrm{pc^{-1}}$)} & \colhead{} & \colhead{} & \colhead{} & \colhead{($\mathrm{km \, s^{-1} \, pc^{-1}}$)} & 
\nocolhead{($\mathrm{km \, s^{-1}}$)} & \colhead{($M_\sun \, \mathrm{yr^{-1}}$)} & \colhead{(Myr)} 
} %
\startdata %
\sidehead{Field-N} %
\cline{1-1} %
F10-E & 105\pm47 & 80\pm30 & 1.625 & 1.2\pm0.7 & 1.0\pm0.6 & 0.5\pm0.3 & 1.21 & (1.3\pm0.9) \times 10^{-4} & 1.8 \\
F60-N & 116\pm52 & 64\pm33 & 1.137 & 0.6\pm0.4 & 0.9\pm0.5 & 0.6\pm0.2 & 1.08 & (1.3\pm0.7) \times 10^{-4} & 1.6 \\
\hline 
\sidehead{Field-S} %
\cline{1-1}
F60-C3 & 84\pm38 & 78\pm26 & 1.529 & 1.4\pm0.8 & 0.9\pm0.7 & 0.4\pm0.2 & 1.16 & (1.0\pm0.7) \times 10^{-4} & 2.7  \\
F60-S & 75\pm34 & 88\pm68 & 1.548 & 1.8\pm1.6 & 1.1\pm0.9 & 0.5\pm0.5 & 0.87 & (0.7\pm0.7) \times 10^{-4} & 1.9 \\
F60-C2 & 78\pm35 & 66\pm44 & 1.201 & 1.0\pm0.8 & 0.8\pm0.5 & 0.3\pm0.4 & 0.28 & (0.2\pm0.3) \times 10^{-4} & 3.6 \\
\enddata %
\tablenotetext{a}{Depletion time if no mass replenishment} %
\end{deluxetable} %

\subsection{Inflow Motion along Filaments} \label{subsec:inflow} %
\begin{figure}[h!] %
\epsscale{0.9} %
\plotone{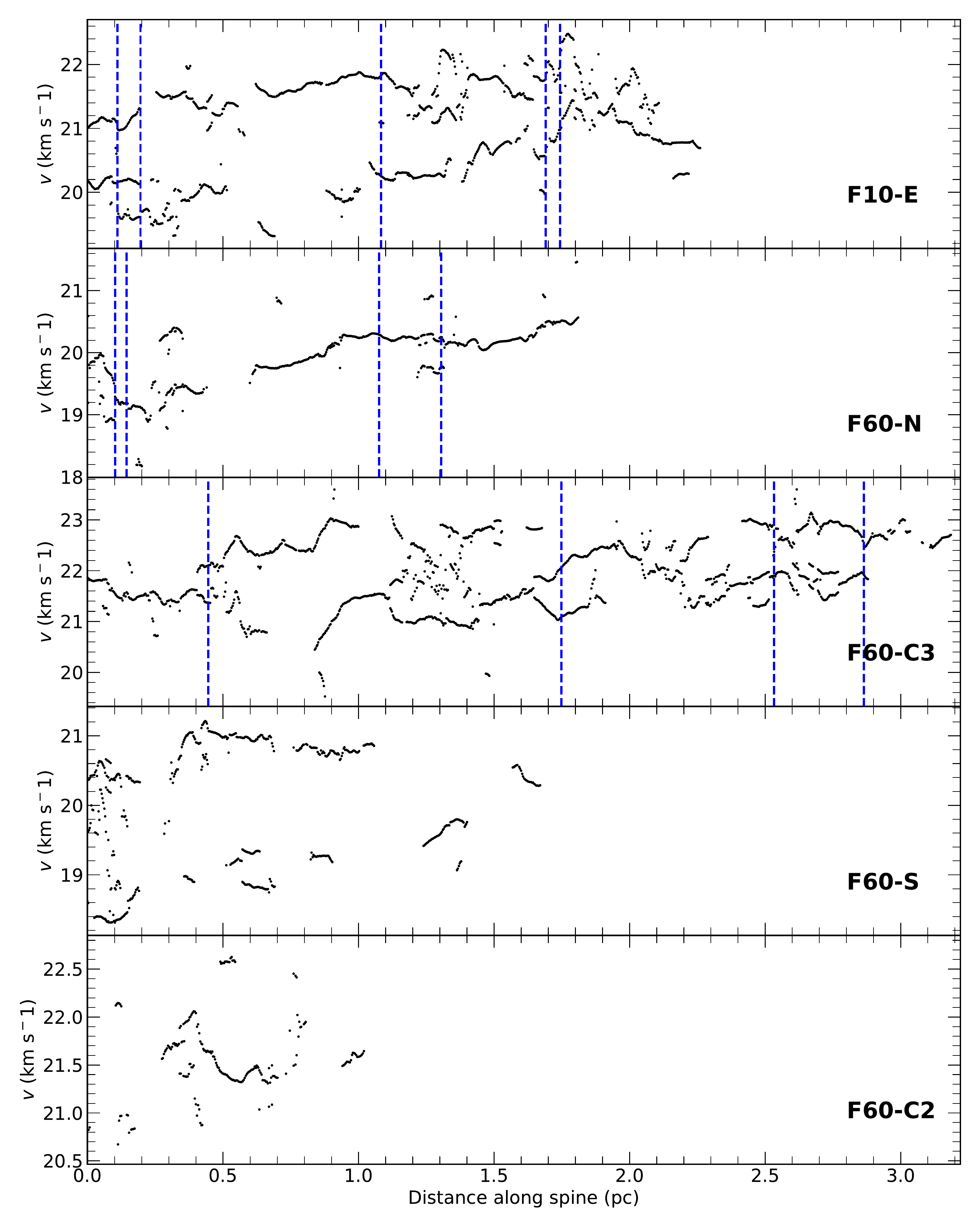} %
\caption{Velocity profiles along filament spines, starting from the end point closer to the hubs.    
The vertical dash lines mark the projected locations of the $3 \; \mathrm{mm}$ dense cores \citep{Ohashi:2016iz} within $0.1 \; \mathrm{pc}$ away from the spines.  
Hub-S and filament F60-C3 are contacted sideway over a region near distance $\sim 1.3 \; \mathrm{pc}$.   
\label{fig:g14_spine_vlsr}} %
\end{figure}

The velocity profiles along individual filament spines are shown in Fig.~\ref{fig:g14_spine_vlsr} with the origin starting at the end point closer to the hubs.
Kinematics along the spines show highly structured filaments with multiple velocity components in many regions. 
Note that kinematics in filament F60-C3 are affected by hub-S (see Fig.~\ref{fig:flmid}b) over a region between projected distance of $1.2-1.4 \; \mathrm{pc}$, where many pixels with four velocity components are present.  
Based on the complex structures in the PPV space  (Fig.~\ref{fig:g14n_ppv} and \ref{fig:g14s_ppv}), simple analyses along the spines will not be able to follow important and relevant structures in the filaments.    
Since we are interested in the collective effect of inflow motions along each filament, we approximate the filament by a cylindrical geometry and perform principal component analysis (PCA) to individual PPV distributions to determine the general orientation and velocity gradient of the filament.   
PCA is a statistical procedure widely used to describe the covariance structures of a set of variables and allows us to identify the principal directions in which the data vary.  
The principal components are found by calculating the eigenvectors and eigenvalues of the covariance matrix.   
The eigenvector with the largest eigenvalue, i.e. the first principal component, is the direction of greatest variation, where the scatter of the data from this axis is minimized.  
The amount of the total variance accounted for by the first principal component is assessed by its significance, which is equal to the percentage of the largest eigenvalue in the sum of all the eigenvalues.  
In our case, the first principal component is used to identify the general orientation of a filament and the velocity gradient along such axis.  
This is an approach similar to previous studies using linear regression, which requires velocity as a function of position, i.e. single velocity in one position  \citep[e.g.][]{Kirk:2013gq,Henshaw:2014ib}.  
Since multiple velocity components are present in our filaments, PCA allows a fit for general orientation and velocity gradient with the full ensemble of PPV components.  
The derived velocity gradient is in the range of $0.3 - 0.6 \; \mathrm{km \, s^{-1} \, pc^{-1}}$ with a mean value of $0.5 \; \mathrm{km \, s^{-1} \, pc^{-1}}$ (Table~\ref{tab:flm}).  
The first principal component has significance higher than 96.6\% in all filaments.  

\subsection{Transonic Turbulent Motions} \label{subsec:lw} %
\begin{figure}[h!] %
\epsscale{0.7} %
\plotone{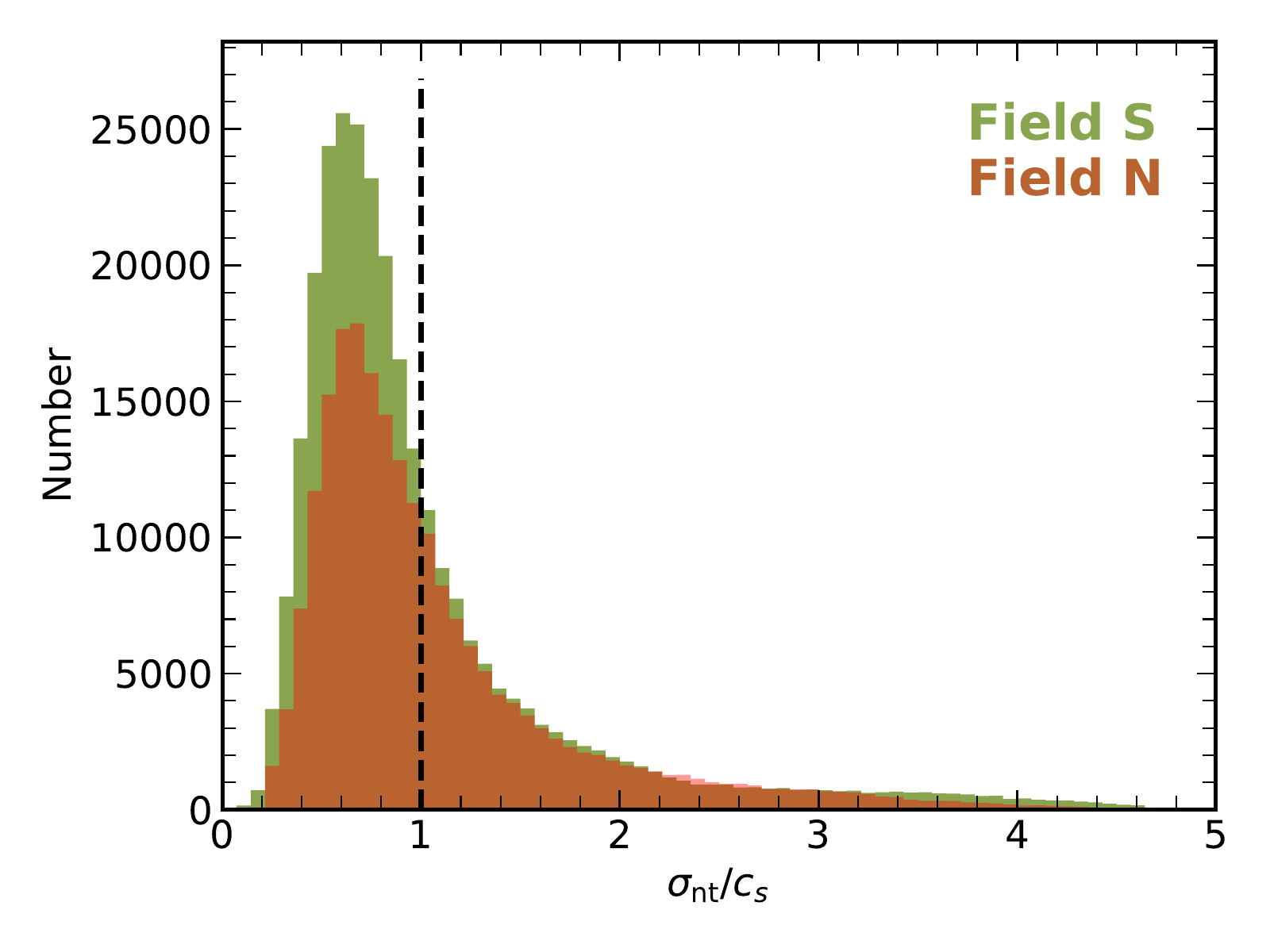} %
\caption{Histogram of all the positions as a function of non-thermal velocity dispersion normalized to local sound speed, $\sigma_\mathrm{nt}/c_s$.  The two fields show a similar distribution with a main peak occurs at $\sigma_\mathrm{nt}/c_s \sim 0.7$ and a slow decay into transonic regime. 
The majority of the positions have non-thermal motions in the subsonic to transonic regimes.  
\label{fig:lw}} %
\end{figure}

In general, the line width of the $\mathrm{N_2H^+} \; (1-0)$ emission with higher angular resolution is narrower than that of the $\mathrm{NH_3} \; (1,1)$ line.  
Since $\mathrm{N_2H^+}$ tends to trace regions of higher densities, the inner part of the filaments may be less affected by radial collapse or turbulence of ambient gas onto filaments. 
We compute the non-thermal velocity dispersion, $\sigma_\mathrm{nt}$, with 
\begin{equation} %
\sigma_\mathrm{nt} = \sqrt{\frac{(\Delta \upsilon)^2}{8 \ln 2} - \frac{k_\mathrm{B} T_g}{m_\mathrm{N_2H^+}} }, 
\label{eq:cnt}
\end{equation} %
where $\Delta \upsilon$ is the observed line width, $k_B$ the Boltzmann constant, $T_g$ the gas temperature, $m_\mathrm{N_2H^+}$ is the mass of $\mathrm{N_2H^+}$.
Since $\mathrm{N_2H^+}$ is of relatively high molecular weight, it is a good tracer to probe the non-thermal velocity dispersion.   
The sound speed in the gas is given by
\begin{equation} %
c_s (T_g) = \sqrt{\frac{k_\mathrm{B} T_g}{\mu m_\mathrm{H}}}, 
\label{eq:cs}
\end{equation} %
where $\mu = 2.33$ is the mean molecular weight, and $m_\mathrm{H}$ is the mass of $\mathrm{H}$.  
The ratio $\sigma_\mathrm{nt}/c_s$ determines whether the gas motion is subsonic with $\sigma_\mathrm{nt}/c_s \le 1$, transonic with $1 < \sigma_\mathrm{nt}/c_s \le 3$, or supersonic with $\sigma_\mathrm{nt}/c_s > 3$.  
This scheme is similar to the choice of transonic regime used by \citet{Arzoumanian:2013gp}.  
Figure~\ref{fig:lw} shows the distribution of the ratio $\sigma_\mathrm{nt}/c_s$ for the entire mosaic regions.   
The two mosaic fields have similar distributions with a peak at $\sigma_\mathrm{nt}/c_s \sim 0.7$ and a slow decay into transonic regime.
Among all the positions with successful fits, 60\%\ have subsonic non-thermal motions and 96\%\ include subsonic and transonic motions altogether. 
Regions with supersonic motions account for only 4\% of the whole population and are preferentially associated with the two hubs, which are main sites of active star formation.  
Still, the limited spectral resolution of $0.2 \; \mathrm{km \, s^{-1}}$ may cause confusion in the fitting algorithm to misidentify two subsonic components very close in velocity as one transonic component.  
The fraction of subsonic velocity components may be further increased if both angular and spectral resolutions improve in future studies. 
Within $0.1 \; \mathrm{pc}$ along filament spines, $\sigma_\mathrm{nt}/c_s$ is in the range of $0.8 - 1.1$ with a mean value of $0.9$, implying moderate transonic non-thermal motions (Table~\ref{tab:flm}).  

\section{Discussion} \label{sec:discussion} %
\subsection{Mass Accretion Rates} %
Once the observed filament length, $\ell_\mathrm{obs}$, and the observed velocity gradient, $\nabla \upsilon_\mathrm{obs}$, are measured, one can  estimate the mass accretion rate by approximating a filament with a cylindrical geometry.   
The observed filament length and velocity gradient are both affected by the projection effect.  
We assume that a filament of mass $M$ has an inclination angle, $i$, with respect to the line of sight.  
Following the method used by \citet{Kirk:2013gq}, one may express the observed quantities with the actual filament length, $\ell$, and flow velocity $\upsilon$ along the filament as   
\begin{eqnarray} %
\ell_\mathrm{obs} &=& \ell \sin i \\
\upsilon_\mathrm{obs} &=& \upsilon \cos i,  
\end{eqnarray} %
where $\upsilon_\mathrm{obs} = \nabla \upsilon_\mathrm{obs} \cdot \ell_\mathrm{obs}$.  
The mass accretion rate, $\dot{M}$, is then given by 
\begin{equation} %
\dot{M} = M \, \nabla \upsilon_\mathrm{obs} \, \tan i = \dot{M}_\mathrm{obs} \tan i.  
\end{equation} %
In practice, it is impossible to identify a filament if it is inclined along the line of sight ($i \sim 0^\circ$) since it gives a null projected length.    
Observational bias is expected in estimates of mass accretion rate due to the projection effect. 
For example, a filament perfectly inclined on the plane of sky ($i \sim 90^\circ$) will not have any detectable velocity gradient. 
Assuming a moderate inclination angle of $i = 45^\circ$, we list in Table~\ref{tab:flm} the mass accretion rate, $\dot{M}_\mathrm{obs}$, and the depletion time, $\tau$, which is the timescale to exhaust all the mass in the filament by the measured accretion rate without mass replenishment from the surroundings.  
The accretion rates will vary in 73\% if we assume a fluctuation of $\pm 15^\circ$, i.e. $30^\circ$ to $60^\circ$ inclination angles.  
To date, only few observations have successfully detected inflow motion along filaments towards their converging hubs  
\citep[e.g.][]{Peretto:2013kt,Peretto:2014cx,Lee:2013bj,Kirk:2013gq,Lu:2018bn}.   
Overall, the inflow motion along the filaments generates an observed velocity contrast, $\upsilon_\mathrm{obs}$, in the range of $0.4$ to $1.5 \; \mathrm{km \, s^{-1}}$.
The flow velocity measured in IRDC G14.2 is in the range of $0.3$ to $1.2 \; \mathrm{km \, s^{-1}}$, which is comparable to previous studies if considering the variation in inclination.  

Using the column density maps \citep{Lin:2016fe}, we estimate the enclosed gas mass within FWHM to be $116 \; \mathrm{M_\odot}$ for both hub-N and hub-S.
The measured core size (FWHM) is $0.25 \; \mathrm{pc}$ for hub-N and $0.30 \; \mathrm{pc}$ for hub-S.  
Assuming a spherical geometry, we obtain the mean number density of $1.37 \times 10^5 \; \mathrm{cm^{-3}}$ and $7.8 \times 10^4 \; \mathrm{cm^{-3}}$ in hub-N and hub-S, respectively.  
These numbers are consistent with those reported by \citet{Busquet:2016jn}.  
The corresponding free-fall time is $t_\mathrm{ff} = 8.4 \times 10^4 \; \mathrm{yr}$ in hub-N and $1.1 \times 10^5 \; \mathrm{yr}$ in hub-S.   
The total mass accretion rate through the two filaments, F10-E and F60-N, connecting to hub-N is $2.6 \times 10^{-4} \; M_\odot \, \mathrm{yr^{-1}}$, which accumulates $22 \; M_\odot$ within one $t_\mathrm{ff}$, about 20\% of the mass in hub-N.  
Similarly, the two filaments, F60-C3 and F60-S, associated with hub-S deliver at a mass accretion rate of $1.7 \times 10^{-4} \; M_\odot \, \mathrm{yr^{-1}}$, which gathers $19 \; M_\odot$ within one $t_\mathrm{ff}$, roughly 16\% of the mass in hub-S.  
Since IRDC~G14.2 is magnetized with mean field strength of $0.35 - 0.55 \; \mathrm{mG}$, the contraction time is likely on a timescale 2--3 times longer than what is expected from a free-fall collapse \citep{Santos:2016cg}. 
Therefore, the filamentary accretion flow may account for nearly half of the mass in the hubs within one  contraction time scale and is sufficient to alter the dynamical evolution of the hubs.  

\subsection{The Virial Parameter in $\mathrm{N_2H^+}$ Filaments} %
The dynamical stability of a system is often assessed by the virial parameter, $\alpha_\mathrm{vir}$.  In the case of filaments, one compares the linear virial mass, $M_\ell^\mathrm{vir}$, to the observed linear mass, $M_\ell$.
The gravitational instability of a pressure-confined isothermal gas layer with uniform magnetic fields has been studied by \citet{Nagai:1998kw}.  
In their models, the layer fragments into filaments, and a subsequent fragmentation to cores may occur in a filament if its linear mass is over a critical value $M_{\ell,\mathrm{crit}} = 2 c_s^2/ G$.  
This model, however, does not include the non-thermal pressure support, which is important in massive star forming regions.  
Following earlier studies \citep[e.g.][]{Fiege:2000ds,Arzoumanian:2013gp}, we apply the effective sound speed, $c_{s,\mathrm{eff}} = \sqrt{c_s^2 + \sigma_\mathrm{nt}^2}$, which combines the local thermal motions of interstellar molecules given by Eq.~(\ref{eq:cs}) with non-thermal motions of the bulk of gas given by Eq.~(\ref{eq:cnt}).   
We estimate the virial mass per unit length with the mean effective sound speed, $\overline{c_{s,\mathrm{eff}}}$, of all the velocity components in a filament
\begin{equation} %
M_{\ell,\mathrm{vir}} = \frac{2 \, \overline{c_{s,\mathrm{eff}}}^2}{G}  \simeq 466 \left( \frac{ \overline{c_{s,\mathrm{eff}}} }{1 \; \mathrm{km \, s^{-1}}} \right)^2  \; M_\odot \, \mathrm{pc}^{-1}.   
\label{eq:Mlvir}  
\end{equation} %
The linear mass, $M_\ell$, however, is calculated from column density integrated along line of sight without substructure details.  
To compare with the virial mass, we compute the mean linear mass accounted for substructures, 
\begin{equation} %
M_\ell^\mathrm{sub} = \frac{M_\ell}{N_\mathrm{sub}},
\end{equation} %
where $N_\mathrm{sub}$ is the average number of substructures in the filament.  
We estimate this average number with $N_\mathrm{sub} = N_\mathrm{vel} / N_\mathrm{pix}$, where $N_\mathrm{vel}$ and $N_\mathrm{pix}$ are the number of velocity components and the number of pixels in a filament, respectively. 
We then compute the virial parameter by comparing the virial mass to the mean linear mass accounted for substructures, 
\begin{equation} %
\alpha_\mathrm{vir} = \frac{M_{\ell,\mathrm{vir}}}{M_\ell^\mathrm{sub}} = \frac{M_{\ell,\mathrm{vir}}}{M_\ell/N_\mathrm{sub}}.      
\end{equation} %
The results are listed in Table~\ref{tab:flm}.  
The value of $M_{\ell,\mathrm{vir}}$ given by Eq.~(\ref{eq:Mlvir}) does not account for relative motion among substructures, whose contribution to pressure support remains unknown. 
Theoretical studies may provide useful insight.        
In addition, only the projected component of relative motion is observable.   
If substructures were to generate additional pressure support, the reported values of $M_{\ell,\mathrm{vir}}$ would be lower limits.  
The value of $\alpha_\mathrm{vir}$ in the $\mathrm{N_2H^+}$ filaments is in the range of $0.6 -1.8$ with a mean value of $1.2$, marginally virialized and likely to be in equilibrium.  
Our measured $\alpha_\mathrm{vir}$ range agrees well with the value range $\alpha_\mathrm{vir} = 0.7 - 2.0$ obtained in the Orion Integral Filament \citep{Hacar:2018eh}.  
Lower values  of $\alpha_\mathrm{vir}$ are commonly observed in regions of high-mass star formation as reported by \citet{Kauffmann:2013bi} using a large complied sample of cloud fragments.

\subsection{Comparison of Properties in Cores, Clumps, and Filaments}  %
\begin{figure}[h!] %
\epsscale{0.8} %
\plotone{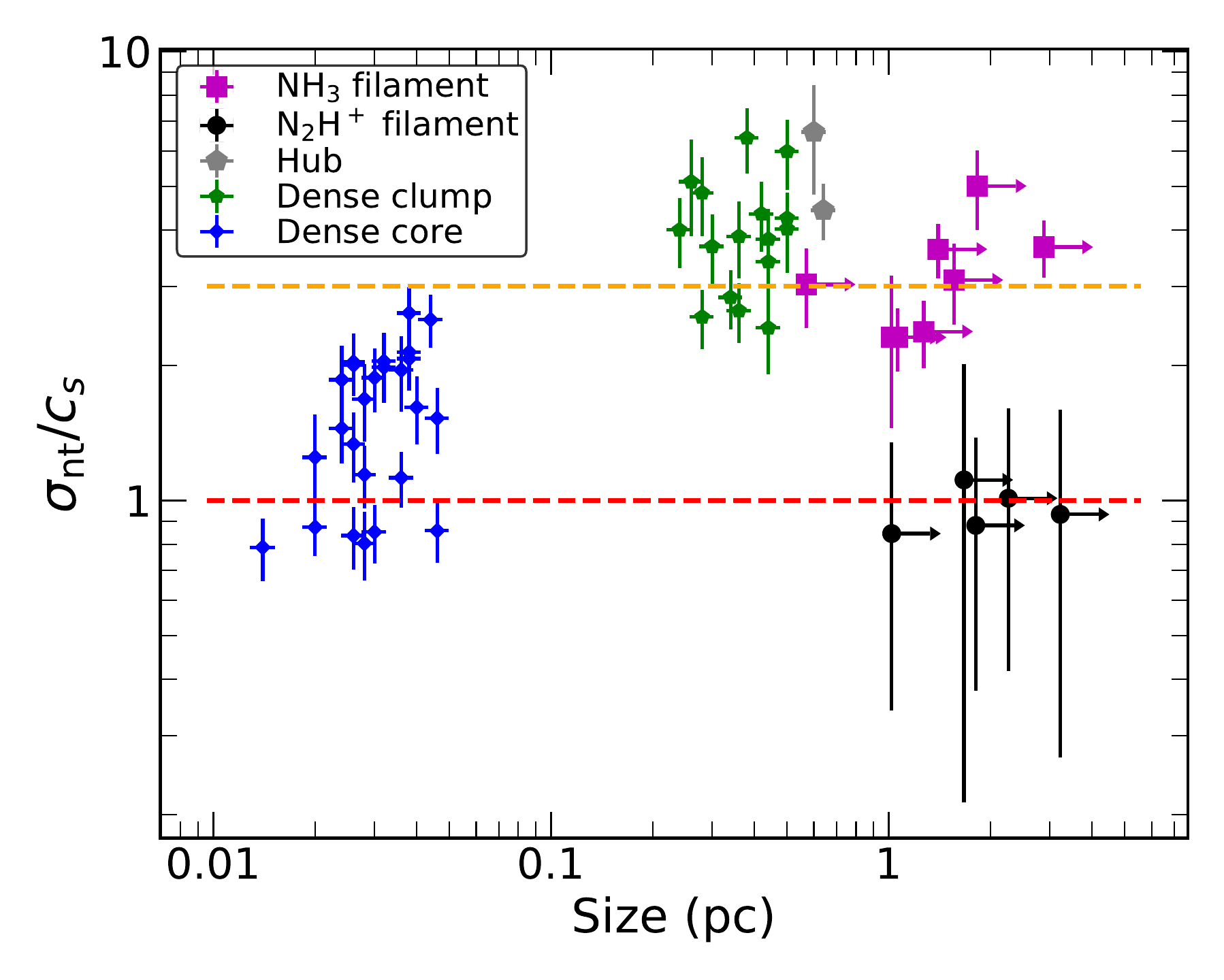} %
\caption{Non-thermal velocity dispersion normalized to local sound speed, $\sigma_\mathrm{nt}/c_s$, versus size of cores/clumps and length of filaments in the compiled sample including cores, clumps, and filaments.    
Black dots indicate measurements of the $\mathrm{N_2H^+}$ filaments in this work.  The data for the $\mathrm{NH_3}$ filaments \citep[magenta squares;][]{Busquet:2013ko}, the $870 \; \mu\mathrm{m}$ continuum  hubs and clumps \citep[gray and green pentagons, respectively;][]{Ohashi:2016iz}, and the $3$mm continuum cores \citep[blue diamonds;][]{Ohashi:2016iz} are also shown.  
The upper limits of the subsonic ($\sigma_\mathrm{nt}/c_s = 1$) and transonic ($\sigma_\mathrm{nt}/c_s = 3$) regimes are indicated by red and orange dashed lines, respectively.     
The $\mathrm{NH_3}$ filaments show supersonic non-thermal motions while the inner part of filaments traced by $\mathrm{N_2H^+}$ are mildly transonic. 
Dense clumps have stronger non-thermal motions than dense cores.  
\label{fig:cnt}} %
\end{figure}

A comparison of the measured quantities in this work, such as the non-thermal velocity dispersion normalized to local sound speed, $\sigma_\mathrm{nt}/c_s$, and the virial parameter, $\alpha_\mathrm{vir}$, is made with those reported in previous studies (Fig.~\ref{fig:cnt}), including filaments identified in the $\mathrm{NH_3}$ emission\footnote{The $\mathrm{NH_3} \; (1,1)$ data have lower spatial resolution ($8\farcs2 \times 7\farcs0$) and spectral resolution ($0.6 \; \mathrm{km \, s^{-1}}$) than do the $\mathrm{N_2H^+} \; (1-0)$ data. 
We set $N_\mathrm{sub} = 1$ for the observed $\mathrm{NH_3}$ filaments, which do not differentiate substructures.  
} %
 \citep[magenta squares;][]{Busquet:2013ko}, hubs and dense clumps identified in the $870 \; \mu\mathrm{m}$ continuum emission \citep[gray and green pentagons, respectively;][]{Ohashi:2016iz}, and dense cores identified in the $3 \, \mathrm{mm}$ continuum emission \citep[blue diamonds;][]{Ohashi:2016iz}. 
In this compiled sample, the size/length of the objects spans a range from $0.007$ to $3.22 \; \mathrm{pc}$.  
Here we consider the observed length of all the filaments to be lower limits of their actual lengths due to projection effects. %
The scale decreases from the $\mathrm{NH_3}$ and $\mathrm{N_2H^+}$ filaments, to the dense clumps, and then to the dense cores.  
We have also revised the measurements for the $\mathrm{NH_3}$ filaments \citep{Busquet:2013ko} using the updated distance of $1.98 \; \mathrm{kpc}$.  

Overall, the $\mathrm{N_2H^+} \; (1-0)$ emissions in filaments show moderate transonic non-thermal gas motions (black dots, Fig.~\ref{fig:cnt}) similar to what has been observed in other filaments \citep[e.g.][]{Hacar:2013tq,Hacar:2018eh,Lu:2018bn}.  
The non-thermal motions observed in the dense clumps and $\mathrm{NH_3}$ filaments tend to be supersonic with $\sigma_\mathrm{nt}/c_s \gtrsim 3$ while subsonic/transonic non-thermal motions ($\sigma_\mathrm{nt}/c_s \sim 1$) are found in the dense cores and $\mathrm{N_2H^+}$ filaments (Fig.~\ref{fig:cnt}).    
Comparing to the mostly supersonic $\mathrm{NH_3}$ emission, the inner volume of the filaments with higher gas density shows weaker non-thermal motions, i.e. smaller $\sigma_\mathrm{nt}$.  
Transonic filaments have been observed and are likely to be dynamically decoupled from the large-scale turbulent fields \citep{Hacar:2013tq,Hacar:2016gx,Chen:2016hj}. 
Meanwhile, the compact $3 \, \mathrm{mm}$ dense cores also have weaker non-thermal motions when comparing to the more extended dense clumps, a phenomenon that has also been reported by \citet{Sanhueza:2017bd} and \citet{Lu:2018bn}.  
There are a few implications of this weaker non-thermal support towards smaller scales.   
Naively, a virialized system under self-gravitation is expected to show an increasing pressure support near the central region, which is the opposite from the trend in $\sigma_\mathrm{nt}/c_s$ in our sample.  
In the case of spherical geometry (clumps and cores), such reduced non-thermal support at small scales may be a feature of global hierarchical collapse \citep{NaranjoRomero:2015ii} or an outcome of dissipation of turbulence giving a transition to coherence  \citep{Goodman:1998dn,Goodman:2009dp,Pineda:2010jq,Gong:2011fb,Chen:2018vm}. 
Nevertheless, an increasing support of magnetic fields that compensates the non-thermal pressure also cannot be ruled out \citep{Kauffmann:2013bi}.     
Numerical simulations of prestellar cores based on the hierarchical collapse scenario develop structures not in hydrostatic equilibrium but with smaller infall velocities in the inner part, giving a smaller $\sigma_\mathrm{nt}/c_s$ \citep{NaranjoRomero:2015ii}.  
The largest velocities occur in the outer parts of the core, making the collapse outside-in.     
On the other hand, dissipation of turbulence due to a reduction of field-neutral coupling at higher density can also produce weaker non-thermal motions at small scales, rendering a pressure difference that may initiate a pressure-driven inflow to allow mass accretion towards the central part of the system \citep{Myers:1998ee}.  
Whether these mechanisms also operate in filaments will need more theoretical  investigation.   

\begin{figure}[h!] %
\epsscale{0.9} %
\plotone{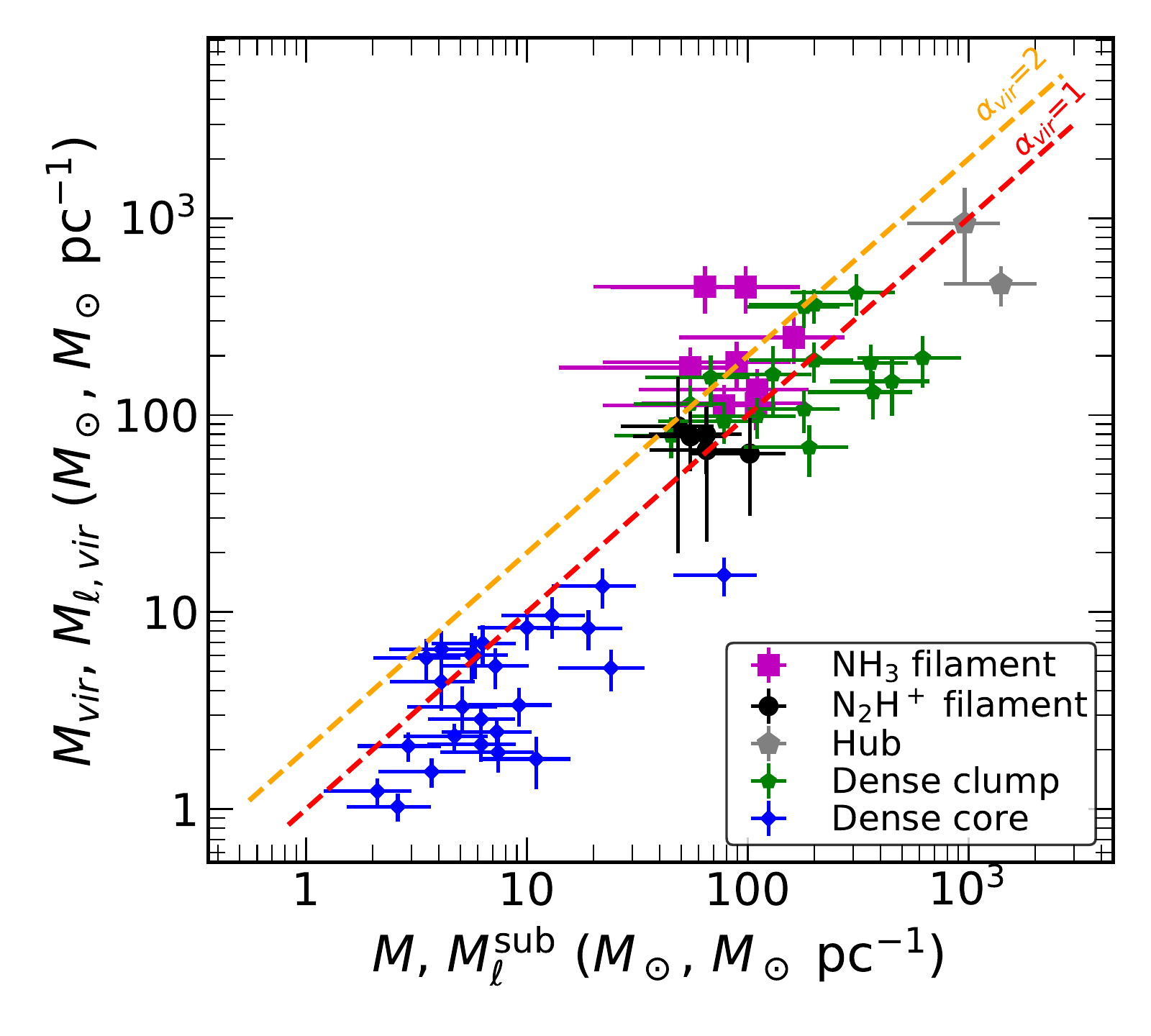} %
\caption{Comparison between the virial mass, $M_\mathrm{vir}$, for cores/clumps and linear virial mass, $M_{\ell,\mathrm{vir}}$, for filaments with the observed gas mass, $M$, or mean linear mass, $M_\ell^\mathrm{sub}$, in IRDC~G14.2.  
The ratio of virial mass to gas mass gives the virial parameter, $\alpha_\mathrm{vir}$.  
The slope corresponding to $\alpha_\mathrm{vir} = 1$ (red dash line) and $\alpha_\mathrm{vir} = 2$ (orange dash line) are also shown.  
Objects with $\alpha_\mathrm{vir} < 1$ are gravitationally bound.     
\label{fig:mvir}} %
\end{figure}

In Fig.~\ref{fig:mvir}, we compare the gas mass, $M$, to the virial mass, $M_\mathrm{vir}$, in cores and clumps \citep{Ohashi:2016iz} as well as the mean linear  mass, $M_\ell^\mathrm{sub}$, to the linear virial mass, $M_{\ell,\mathrm{vir}}$, in filaments \citep[this work and][]{Busquet:2013ko}.  
Sources appear below the boundary line of $\alpha_\mathrm{vir} = 1$ (red dash line) are expected to be gravitationally bound.  
In our sample, the $\mathrm{NH_3}$ filaments have significantly higher linear virial mass due to their supersonic nature.  
The non-thermal motions in the $\mathrm{N_2H^+}$ filaments are mildly transonic.  
Similar to the $\mathrm{N_2H^+}$ filaments, the dense clumps are also in equilibrium.  
Dense cores, the smallest scales in our sample, are gravitationally bound with significantly lower values of $\alpha_\mathrm{vir} < 1$.

\subsection{Uncertainties in Mass Estimates} %
A few factors may contribute to uncertainties in the mass estimates in our sample.  
We calculate uncertainties for derived quantities by propagating errors in dependent variables.  
The uncertainty in the distance measurement of IRDC~G14.2 is roughly 7\% \citep{Xu:2011da}, which affects all the quantities involving physical scales and masses derived from flux measurements.    
Following the discussion by \citet{Sanhueza:2017bd}, we adopt uncertainties of 28\% and 23\% for the dust opacity and gas-to dust mass ratio.  
The uncertainties in the dust temperature and column density measurements with the iterative SED fits \citep{Lin:2016fe} are around 20\% and 10\%, respectively.  
Hence, the uncertainty in the gas linear mass after propagating all the errors is roughly 45\%.  
The uncertainties in the virial mass estimates arise from the dispersion in the effective sound speed, $c_{s,\mathrm{eff}}$, and are listed in Table~\ref{tab:flm}.       
The uncertainty in the ammonia temperature measurements applied to the $\mathrm{NH_3}$ filaments, dense clumps, and dense cores are assumed to be $3 \; \mathrm{K}$, which is a more conservative estimate.      
The absolute flux measurements of ALMA Band~3 is good within 5\% (Section~\ref{sec:obs}) so the typical uncertainty in mass estimates of dense cores is around 48\%.  
Regarding mass estimates for $\mathrm{NH_3}$ filaments, the uncertainty in the ammonia abundance, $X_\mathrm{NH_3}$, is assumed to be 67\% based on the dispersion in the $\mathrm{NH_3}$ abundance studies in infrared dark clouds \citep{Pillai:2006ku}. 
In the current analyses, we ignore uncertainties caused by plausible biases in  identifying cores, clumps, and filaments.  
Such uncertainty may require simulations to obtain a fair estimate.  

\subsection{Dynamical Stability and Virial Parameters} %
\begin{figure}[h!] %
\epsscale{1.2} %
\plotone{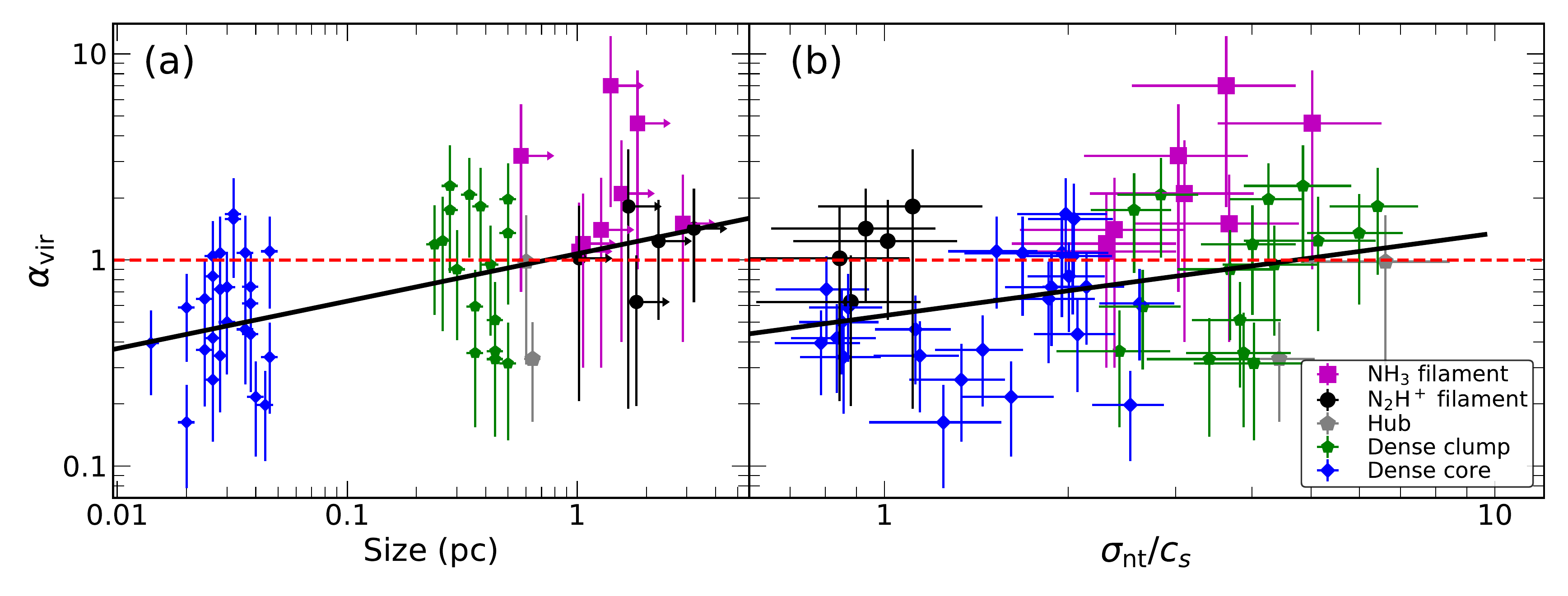} %
\caption{(a) The virial parameter, $\alpha_\mathrm{vir}$, as a function of the size, $s$, of the observed objects.  
Red dash line indicates $\alpha_\mathrm{vir} = 1$.  
The decreasing trend in $\alpha_\mathrm{vir}$ with decreasing scale persists as previously reported by \citet{Ohashi:2016iz}. 
Linear regression derived from the bootstrapping statistical samples gives $\log \alpha_\mathrm{vir} = (0.23 \pm 0.06) \log s + (0.03 \pm 0.07)$ (black line).  
(b) The virial parameter, $\alpha_\mathrm{vir}$, as a function of non-thermal velocity dispersion normalized to local sound speed, $\sigma_\mathrm{nt}/c_s$.   
Filaments show slightly higher $\alpha_\mathrm{vir}$ than cores and clumps. 
A decreasing trend in $\alpha_\mathrm{vir}$ with decreasing $\sigma_\mathrm{nt}/c_s$ is seen, suggesting that objects with smaller $\sigma_\mathrm{nt}/c_s$ tend to be more dominated by gravity. 
Linear regression renders $\log \alpha_\mathrm{vir} = (0.4 \pm 0.2) \log (\sigma_\mathrm{nt}/c_s) - (0.27 \pm 0.07)$ (black line).  
\label{fig:alpha}} %
\end{figure}

We further investigate the dependence in the virial parameter, $\alpha_\mathrm{vir}$, with physical scales, $s$, and non-thermal velocity dispersion normalized to local sound speed, $\sigma_\mathrm{nt}/c_s$.  
A decreasing trend in $\alpha_\mathrm{vir}$ with decreasing scales, from $\mathrm{NH_3}$ filaments to $\mathrm{N_2H^+}$ filaments, then to dense clumps, and down to dense cores, can be discerned (Fig.~\ref{fig:alpha}a).    
To investigate whether this decreasing trend is robust, we perform linear regression for a bootstrapping statistical sample of 50,000 synthetic data sets.  
Care has been taken to assure sufficient sample size for convergence.  
We then calculate the mean and standard deviation of the slopes and intercepts derived from the sample.  
This analysis is necessary due to the unknown inclination angle of the filaments.
For cores and clumps, we assume normal distribution of uncertainty in physical scale (diameter), $s$, and virial parameter, $\alpha_\mathrm{vir}$.    
In the case of filaments, the uncertainty in $\alpha_\mathrm{vir}$ is assumed to be a normal distribution. 
Because of the unknown intrinsic length of a filament, we assume a uniform distribution of inclination angles and allow the deprojected length to reach a given maximum length, $\ell_\mathrm{max}$.  
We vary $\ell_\mathrm{max}$ to examine how the slope and intercept depend on $\ell_\mathrm{max}$.   
Our test finds a weak dependence that $\log \alpha_\mathrm{vir} = (0.23 \pm 0.06) \log s + (0.03 \pm 0.07)$ for maximum intrinsic filament length in the range of $\ell_\mathrm{max} = 5 - 8 \; \mathrm{pc}$ (black line in Fig.~\ref{fig:alpha}a).  
The slope decreases monotonically to 0.22 for longer intrinsic filament lengths up to $20 \; \mathrm{pc}$.  
The $\alpha_\mathrm{vir}$ also shows a decreasing trend with $\sigma_\mathrm{nt}/c_s$.  
We perform similar bootstrapping statistical analysis with a normal distribution for all the uncertainties and obtain 
$\log \alpha_\mathrm{vir} = (0.4 \pm 0.2) \log (\sigma_\mathrm{nt}/c_s) - (0.27 \pm 0.07)$ (black line in Fig.~\ref{fig:alpha}b).  

The $\mathrm{N_2H^+}$ filaments added in this work are consistent with such behavior in $\alpha_\mathrm{vir}$ previously reported by \citet[][Fig.~\ref{fig:alpha}a]{Ohashi:2016iz}.    
Since lower $\alpha_\mathrm{vir}$ indicates a condition of gravity dominating over the pressure support, this decreasing trend suggests an increasingly important role of gravity at small scales.  
Meanwhile, $\alpha_\mathrm{vir}$ also shows a general decreasing trend with decreasing non-thermal motions, $\sigma_\mathrm{nt}/c_s$ (Fig.~\ref{fig:alpha}b), suggesting that objects with smaller $\sigma_\mathrm{nt}/c_s$ tend to be more dominated by gravity.  
As also shown in Fig.~\ref{fig:cnt}, the non-thermal motions are supersonic in the $\mathrm{NH_3}$ filaments and dense clumps, transonic in the $\mathrm{N_2H^+}$ filaments and dense cores.  
In general, filaments show a slightly higher $\alpha_\mathrm{vir}$ comparing to dense clumps and cores.  
The $\mathrm{N_2H^+}$ filaments and dense clumps have $\alpha_\mathrm{vir} \sim 1$, likely to be in equilibrium, while dense cores, though transonic, are gravitationally bound with $\alpha_\mathrm{vir} < 1$.   

Our current analyses of $\alpha_\mathrm{vir}$ in the cores and clumps do not include magnetic fields, which is expected to provide additional support against self-gravity \citep{VanLoo:2014kx} and increase the value of $\alpha_\mathrm{vir}$ . 
The magnetic field strength in IRDC G14.2 reported by \citet{Santos:2016cg} is in the range $0.32$--$0.55 \; \mathrm{mG}$, corresponding to the Alfv\'{e}n Mach number in the range of ${\cal M}_A = 0.5$--$0.8$.  
For a magnetized cloud with uniform density distribution, the virial mass is given by \citep{Lu:2015do}  
\begin{equation} %
M_{\mathrm{vir}, B} = \frac{5 R \, c_{s,\mathrm{eff}}^2}{G} \left( 1 + \frac{1}{2 {\cal M}_A^2} \right) = M_\mathrm{vir} \left( 1 + \frac{1}{2 {\cal M}_A^2} \right),   
\end{equation} %
where $M_\mathrm{vir} = 5 R \, c_{s,\mathrm{eff}}^2/G$ is the virial mass for a non-magnetized sphere.   
Hence the magnetic fields in IRDC~G14.2 are able to increase the non-magnetized virial mass, $M_\mathrm{vir}$, by a factor of $1.8$--$3.0$, which is not necessarily negligible.  
However, since the field strength was measured in much larger scale, it is not clear how $\alpha_\mathrm{vir}$ will vary at scales of our filaments and cores.
Meanwhile, an inflow toward a hub will drain the mass in a filament unless mass replenishment occurs to sustain the filament.  
If filaments are long-lasting features, mass replenishment is needed and will most likely come from the surroundings.  
Striation features around filaments are thought to be related to this accretion scenario \citep[e.g.][]{Palmeirim:2013da}.  
Observationally, such a radial collapse has only been reported in a filament in Serpens South \citep{Kirk:2013gq}.
The velocity gradients along filaments observed in IRDC~G14.2 may drain the filaments and induce mass replenishment from the surroundings, perhaps a radial accretion onto the filaments, producing external ram pressure that helps to keep the filaments bound.    

\subsection{Massive Star Formation Scenarios} %
A few theoretical scenarios have been proposed to explain massive star formation.  
In the turbulent core model \citep{McKee:2002vc}, massive stars form via a monolithic collapse of a massive core, which is approximately in hydrostatic equilibrium with pressure support from turbulence.  
Additional feedback from low-mass protostars such as radiative heating also help to suppress fragmentation in massive cores.      
Hence, cores forming massive stars usually harbor one or a few stars \citep{Krumholz:2007hs}.    
Alternatively, scenarios allowing continuous mass accretion through the protostellar phase have also been developed.     
The competitive accretion model \citep{Bonnell:2001fk} describes the accretion in clusters as a dynamical phenomena. 
A cloud first fragment into cores of thermal Jeans mass and form a cluster of low-mass protostars.  
Subsequent Bondi-Hoyle type accretion of surrounding gas in the parent clump allow protostars to grow in mass.  
Those protostars located near the center of the cluster accrete gas of higher densities and gain mass faster, having a better chance to become massive stars.     
Recently, the global hierarchical collapse model \citep{VazquezSemadeni:2009fa,BallesterosParedes:2011gk,Hartmann:2012yq} advocates a picture of molecular clouds in a state of hierarchical and chaotic gravitational collapse, in which local centers of collapse develop throughout the cloud while the cloud itself is contracting.    
The collapse applies to all scales but not necessarily starts at precisely the same instant.
In this scenario, a small number of stars may form early throughout the cloud before global contraction increases the gas density and the bulk of stellar population is formed in the center.   
This model has reproduced quantitatively a few observational properties of star-forming clusters, such as high local star formation rates with low global efficiencies \citep{VazquezSemadeni:2009fa} and the age spreads in young cluster members \citep{Hartmann:2012yq}.  

To date, a good range of surveys have been conducted in the IRDC~G14.2.  
A star-forming scenario that can explain the main observational results is gradually emerging.      
Young stellar populations observed in the X-ray and infrared wavebands reveal a significant deficit of high-mass YSOs \citep{Povich:2009dm,Povich:2010iv,Povich:2016jm}.  
This absence of massive stars in the intermediate stage of cluster formation has been reproduced in simulation involving global hierarchical collapse \citep{VazquezSemadeni:2017gq}.  
In the millimeter waveband, clumps and cores have been identified to study core mass function \citep{Busquet:2016jn,Ohashi:2016iz}.  
Both prestellar and protostellar cores are gravitationally bound with low values of virial parameter $\alpha_\mathrm{vir} < 1$.   
None of prestellar and protostellar cores is more massive than $22 \; M_\odot$, suggesting that cores do not acquire all their mass before forming a protostar but continuously gain mass through protostellar phase. 
In contrast with forming a few protostars via a monolithic collapse of a massive core, the dense cores in IRDC~G14.2 are likely accreting from the surroundings that are fed by their parent clumps or filaments  \citep{Gomez:2014bu}.   
In the current study, we find protostars and cores are preferentially located in the hubs and filaments.   
The filaments deliver mass to the hubs with sufficiently high accretion rates to affect the hub dynamics within one free-fall time ($\sim 10^5 \; \mathrm{yr}$).  
These observational features are consistent with the global hierarchical collapse scenario if IRDC~G14.2 produces massive protostars later in time and matures with the Salpeter IMF. 
 
\subsection{Alternative Dynamical Interpretation and Substructures in Filaments} %
The unknown inclination introduces an unavoidable bias in identifying a filament and a fairly large uncertainty in the estimates of the accretion rate along the axis.  
It has also rendered two plausible scenarios, inflow or expansion, for the observed velocity gradients as discussed in the case of IRDC~G035.39$-$00.33 \citep{Henshaw:2014ib}.   
If filaments are in expansion, higher pressure and stronger non-thermal motions will be expected in hubs and filaments.  
The substructures in the $\mathrm{N_2H^+}$ filaments show subsonic non-thermal motions (Fig.~\ref{fig:lw}).
Both the hubs and $\mathrm{N_2H^+}$ filaments are gravitationally bound (Fig.~\ref{fig:mvir}).    
Hence the expansion scenario seems less favorable in IRDC G14.2.

Recent studies on filaments have shown the presence of sub-structures, i.e. fibers, which collectively form a filament \citep{Li:2012ep,Hacar:2013tq,Hacar:2018eh,Sokolov:2017dz}.  
These sub-structures have also been reproduced in numerical simulations and are important to our understanding of filament formation mechanisms \citep{Smith:2014tx,Moeckel:2015cj,Smith:2016dn,Clarke:2017cm}.  
It is not yet clear whether fibers are long-lived, pre-existing structures or density perturbation developed during accretion from an inhomogeneous turbulent medium.  
The internal kinematics among fibers may also affect the dynamical stability of a filament.  
With the limited sensitivity and velocity resolution of $0.2 \; \mathrm{km \, s^{-1}}$ in our current observations, we notice multiple velocity components present in roughly $1/3$ of our spectra but cannot trace and differentiate individual fibers reliably.  
Although weak emission is detected in the total intensity maps, we were not able to obtain successful fits.  This produces disconnected short segments in one seemingly coherent structure in the PPV space.  
In addition, a small number of velocity components are present between fibers.  
It is not clear whether all these components may be neglected by assuming their spectra resulted from line blending of components in the neighboring fibers. 
This issue occurs in every filament but particularly severe for filaments in Field-S.  
Future observations with improved spatial and spectral resolution will be needed if each individual fibers are to be robustly identified. 

\section{Conclusion} \label{sec:conclusion} %
We have performed full-synthesis imaging to map the $\mathrm{N_2H^+} \; (1-0)$ emission in the IRDC~G14.2 with two mosaic fields that cover hub-N and hub-S as well as their associated filaments using the ALMA 12-m Array, the ACA, and the TP array.   
Our observations resolve the filaments with resolutions of $\sim 0.034 \; \mathrm{pc}$.  
Kinematics are derived from sophisticated spectral fitting algorithm that accounts for line blending, large optical depth, and multiple velocity components.  
Our main findings are as follows: 
\begin{enumerate} %
\item We identify five filaments with the dense, quiescent gas tracer $\mathrm{N_2H^+}$ using the {\sf FilFinder} package. 
Embedded YSOs and dense cores are preferentially associated with hubs and filaments.      
Large-scale velocity gradients are detected, suggestive of accretion flows towards the two dominant hubs, where proto-clusters are located.  
In general, filaments show mildly transonic non-thermal motions, and $\sim 1/3$ of the positions show multiple velocity components.  
\item Principal component analysis (PCA) is used to find the dominant flow direction and velocity gradient in each filament.      
Assuming a moderate inclination angle of $i = 45^\circ$, mass accretion rates along filaments are in the range of  $(0.2 - 1.3) \times 10^{-4} \; M_\odot \, \mathrm{yr^{-1}}$.   
Simple estimates show that the accretion by filaments is significant to affect the dynamics of hubs within one free-fall time ($\sim 10^5 \; \mathrm{yr}$).  
\item  The $\mathrm{N_2H^+}$ emission profiles are analyzed with the {\sf RadFil} package for measuring the width of the filaments.  The width ranges from 0.05 to 0.09 with a mean value of $0.07 \; \mathrm{pc}$, which is smaller than but comparable to the universal $0.1 \; \mathrm{pc}$ width reported by previous {\it Herschel} studies.  
\item Our $\mathrm{N_2H^+}$ filaments are marginally virialized and likely to be in equilibrium with a mean value of $\alpha_\mathrm{vir} \sim 1.2$.  
Magnetic fields may play a role to provide additional support in filaments with small $\alpha_\mathrm{vir}$. 
\item A comparison study of $\alpha_\mathrm{vir}$ measured in the $\mathrm{NH_3}$ filament, $\mathrm{N_2H^+}$ filament, $870 \; \mu\mathrm{m}$ dense clumps, and $3 \; \mathrm{mm}$ dense cores is made.  
The $\mathrm{NH_3}$ filaments and dense clumps show supersonic non-thermal motions while the $\mathrm{N_2H^+}$ filaments and dense cores are mostly subsonic and transonic.  
The decreasing trend in $\alpha_\mathrm{vir}$ with decreasing scales persists, suggesting an increasingly important role of gravity at small scales.  
We also found that $\alpha_\mathrm{vir}$ decreases with decreasing non-thermal motions.  
The large-scale filamentary accretion flows are likely feeding hubs, which harbor dense small-scale structures. 
In combination with the absence of high-mass protostars and massive cores, our observational resutls are consistent with the global hierarchical collapse scenario. 
\end{enumerate} %

\acknowledgments %
We are indebted to a careful anonymous referee, who helped significantly to improve the paper.  
This work is supported by the Taiwan Ministry of Science and Technology, project MOST 106-2119-M-007-022-MY3 and 105-2119-M-007-022-MY3. 
G.B. is supported by the MINECO (Spain) AYA2014-57369-C3 and AYA2017-84390-C2-2-R grants.
P.S. was financially supported by Grant-in-Aid for Scientific Research (KAKENHI Number 18H01259) of
Japan Society for the Promotion of Science (JSPS). 
A.P. acknowledges financial support from UNAM-PAPIIT IN113119 grant, M\'exico.
This paper makes use of the following ALMA data: ADS/JAO.ALMA\#2013.1.00312.S, \#2015.1.00418.S. ALMA is a partnership of ESO (representing its member states), NSF (USA) and NINS (Japan), together with NRC (Canada), MOST and ASIAA (Taiwan), and KASI (Republic of Korea), in cooperation with the Republic of Chile. The Joint ALMA Observatory is operated by ESO, AUI/NRAO and NAOJ.


%

\vspace{5mm} %
\facility{ALMA} %


\software{CASA \citep{McMullin:2007tj},  
          MIRIAD \citep{Sault:1995ub},
          FilFinder \citep{Koch:2015dc},
          RadFil \citep{Zucker:2018eg},  
          TOPCAT \citep{Taylor:2005wq},  
          } %



\appendix

\section{N$_2$H$^+$ $(1-0)$ Velocity Channel Maps \label{sec:chan_maps}} %
\begin{figure} %
\epsscale{1.0} %
\plotone{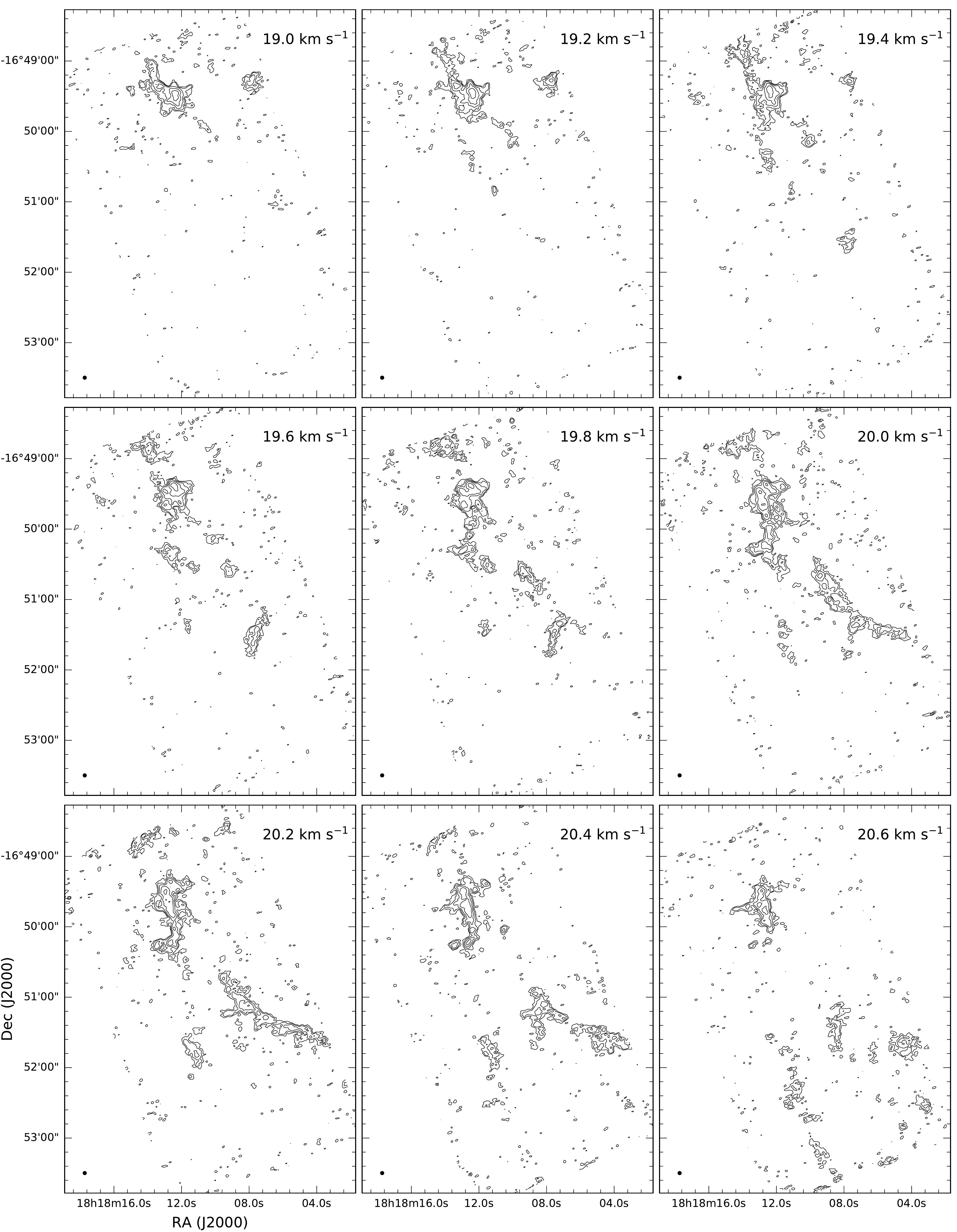} %
\caption{Velocity channel maps in the Field-N. Contour levels are 0.08, 0.13, 0.21, 0.34, 0.54, $0.87 \; \mathrm{Jy \, beam^{-1}}$ with a beam size of $3\farcs48$.  The complete figure set (4 images) is available in the online journal.  \label{fig:g14n_fullchan}} %
\end{figure}

\begin{figure} %
\epsscale{1.0} %
\plotone{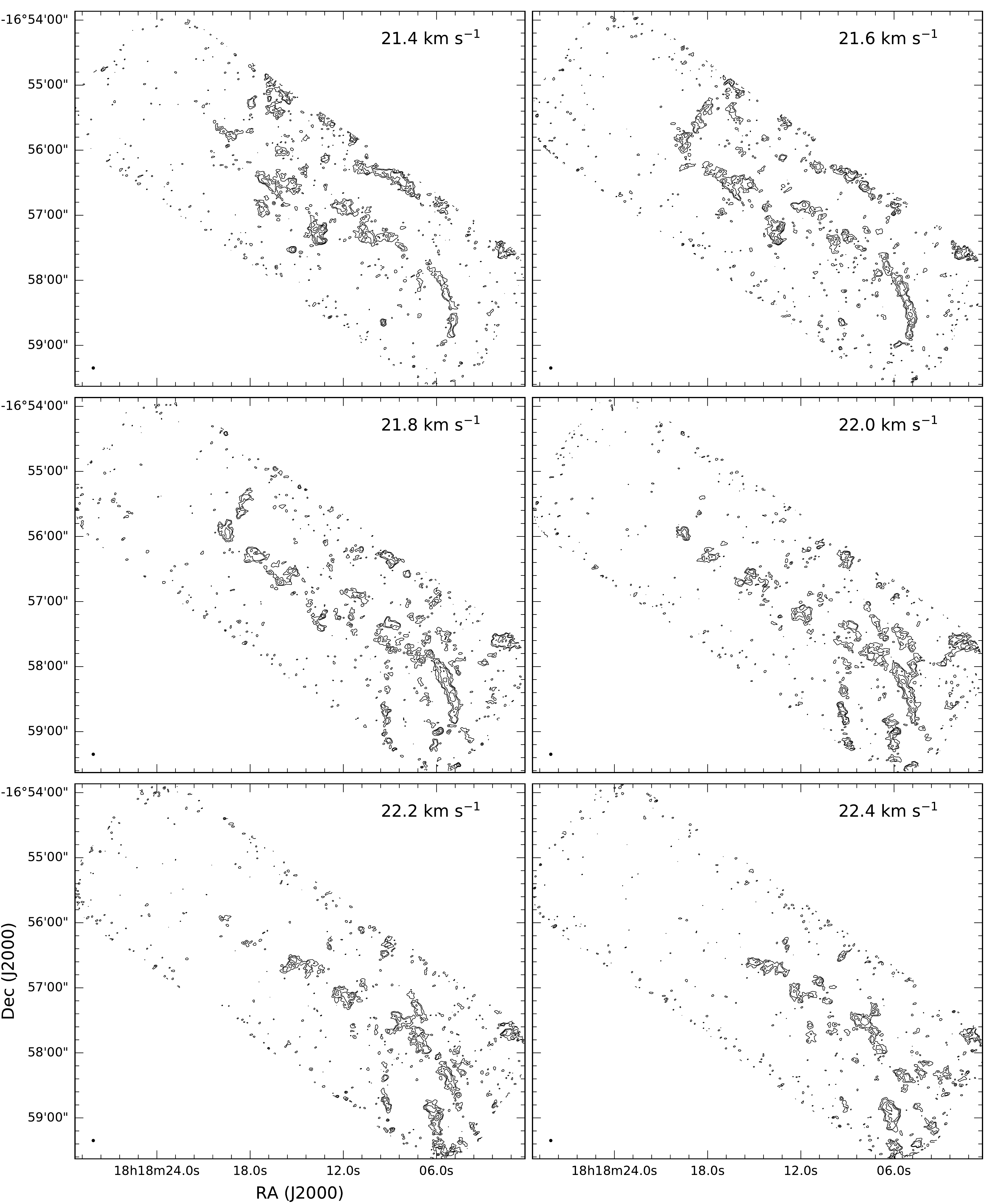} %
\caption{Velocity channel maps in the Field-S. Contour levels are 0.07, 0.11, 0.16, 0.25, 0.39, $0.59 \; \mathrm{Jy \, beam^{-1}}$ with a beam size of $3\farcs15$.  The complete figure set (5 images) is available in the online journal.  \label{fig:g14s_fullchan}} %
\end{figure}

Here we present the velocity channel maps of the isolated $F_1F = 01 \rightarrow 12$ component of the $\mathrm{N_2H^+} \; (1-0)$ emission with velocity step of $0.2 \; \mathrm{km \, s^{-1}}$ for Field-N (Fig.~\ref{fig:g14n_fullchan}) and Field-S (Fig.~\ref{fig:g14s_fullchan}).      

\figsetstart %
\figsetnum{18}
\figsettitle{Velocity channel maps in the Field-N.}

\figsetgrpstart
\figsetgrpnum{18.1}
\figsetgrptitle{Images of Field-N channel 01 to 09}
\figsetplot{f18_1.pdf}
\figsetgrpnote{Velocity channel maps in the Field-N. Contour levels are 0.08, 0.13, 0.21, 0.34, 0.54, $0.87 \; \mathrm{Jy \, beam^{-1}}$ with a beam size of $3\farcs48$. }
\figsetgrpend

\figsetgrpstart
\figsetgrpnum{18.2}
\figsetgrptitle{Images of Field-N channel 10 to 18 }
\figsetplot{f18_2.pdf}
\figsetgrpnote{Velocity channel maps in the Field-N. Contour levels are 0.08, 0.13, 0.21, 0.34, 0.54, $0.87 \; \mathrm{Jy \, beam^{-1}}$ with a beam size of $3\farcs48$.}
\figsetgrpend

\figsetgrpstart
\figsetgrpnum{18.3}
\figsetgrptitle{Images of Field-N channel 19 to 27 }
\figsetplot{f18_3.pdf}
\figsetgrpnote{Velocity channel maps in the Field-N. Contour levels are 0.08, 0.13, 0.21, 0.34, 0.54, $0.87 \; \mathrm{Jy \, beam^{-1}}$ with a beam size of $3\farcs48$.}
\figsetgrpend

\figsetgrpstart
\figsetgrpnum{18.4}
\figsetgrptitle{Images of Field-N channel 28 to 30 }
\figsetplot{f18_4.pdf}
\figsetgrpnote{Velocity channel maps in the Field-N. Contour levels are 0.08, 0.13, 0.21, 0.34, 0.54, $0.87 \; \mathrm{Jy \, beam^{-1}}$ with a beam size of $3\farcs48$.}
\figsetgrpend

\figsetend %

\figsetstart %
\figsetnum{19}
\figsettitle{Velocity channel maps in the Field-S.}

\figsetgrpstart
\figsetgrpnum{19.1}
\figsetgrptitle{Images of Field-S channel 01 to 06}
\figsetplot{f19_1.pdf}
\figsetgrpnote{Velocity channel maps in the Field-S. Contour levels are 0.07, 0.11, 0.16, 0.25, 0.39, $0.59 \; \mathrm{Jy \, beam^{-1}}$ with a beam size of $3\farcs15$.}
\figsetgrpend

\figsetgrpstart
\figsetgrpnum{19.2}
\figsetgrptitle{Images of Field-S channel 07 to 12 }
\figsetplot{f19_2.pdf}
\figsetgrpnote{Velocity channel maps in the Field-S. Contour levels are 0.07, 0.11, 0.16, 0.25, 0.39, $0.59 \; \mathrm{Jy \, beam^{-1}}$ with a beam size of $3\farcs15$.}
\figsetgrpend

\figsetgrpstart
\figsetgrpnum{19.3}
\figsetgrptitle{Images of Field-S channel 13 to 18 }
\figsetplot{f19_3.pdf}
\figsetgrpnote{Velocity channel maps in the Field-S. Contour levels are 0.07, 0.11, 0.16, 0.25, 0.39, $0.59 \; \mathrm{Jy \, beam^{-1}}$ with a beam size of $3\farcs15$.}
\figsetgrpend

\figsetgrpstart
\figsetgrpnum{19.4}
\figsetgrptitle{Images of Field-S channel 19 to 24 }
\figsetplot{f19_4.pdf}
\figsetgrpnote{Velocity channel maps in the Field-S. Contour levels are 0.07, 0.11, 0.16, 0.25, 0.39, $0.59 \; \mathrm{Jy \, beam^{-1}}$ with a beam size of $3\farcs15$.}
\figsetgrpend

\figsetgrpstart
\figsetgrpnum{19.5}
\figsetgrptitle{Images of Field-S channel 25 to 30 }
\figsetplot{f19_5.pdf}
\figsetgrpnote{Velocity channel maps in the Field-S. Contour levels are 0.07, 0.11, 0.16, 0.25, 0.39, $0.59 \; \mathrm{Jy \, beam^{-1}}$ with a beam size of $3\farcs15$.}
\figsetgrpend

\figsetend

\section{N$_2$H$^+$ Spectral Models and the Fitting Procedure \label{sec:hfcfit}} %
The fitting procedure in the current work uses an improved algorithm to handle a large amount of spectral data based on our previous studies \citep{Chen:2010hn,Chen:2011bo}.  
In low temperature environment, the radiation of cosmic microwave background at $T_\mathrm{bg} = 2.7 \; \mathrm{K}$ may not be neglegible.   
One can find the intensity of line emission and continuum emission of a cloud to be  
\begin{eqnarray} %
I_\nu^\mathrm{line} &=& B_\nu(T_\mathrm{bg}) \, e^{-(\tau_\nu^\mathrm{line}+\tau_\nu^\mathrm{cont})} + S_\nu \left[ 1-e^{-(\tau_\nu^\mathrm{line}+\tau_\nu^\mathrm{cont})} \right], \nonumber \\
I_\nu^\mathrm{cont} &=& B_\nu(T_\mathrm{bg}) \, e^{-\tau_\nu^\mathrm{cont}} + S_\nu \left[ 1-e^{-\tau_\nu^\mathrm{cont}} \right], \nonumber 
\end{eqnarray} %
where $B_\nu(T_\mathrm{bg})$ is the Planck function at a temperature $T_\mathrm{bg}$, $S_\nu$ the source function determined by the gas in the cloud, $\tau_\nu^\mathrm{line}$ and $\tau_\nu^\mathrm{cont}$ the respective optical depth of the line and continuum.  
Assuming the gas along line of sight is  isothermal at a gas temperature of $T_g$, one can approximate the source function by $S_\nu = B_\nu(T_g)$.   
In observations, the intensity of line emission is obtained after subtracting the continuum emission 

\begin{eqnarray} %
I_\nu &=& I_\nu^\mathrm{line} - I_\nu^\mathrm{cont} \nonumber \\  
    &=& e^{-\tau_\nu^\mathrm{cont}} \, (1 - e^{-\tau_\nu^\mathrm{line}}) \left[ B_\nu(T_g) - B_\nu(T_\mathrm{bg}) \right].  \nonumber 
\end{eqnarray} %
In terms of brightness temperature, one finds  
\begin{equation} %
T_b(\nu) = e^{-\tau_\nu^\mathrm{cont}} \, (1-e^{-\tau_\nu^\mathrm{line}}) [J_\nu(T_g) - J_\nu(T_\mathrm{bg})], \label{eq:Tb}
\end{equation} %
where $J_\nu (T) \equiv (c^2/2 k \nu^2) B_\nu(T) = (h\nu/k)(e^{h\nu/k T} -1)^{-1}$.  
The emission at a given velocity is described by three parameters, the velocity $\upsilon_i$, the column density $N_i$, and the full-width at half-maximum (FWHM) as line width $\Delta \upsilon_i$.  
The optical depth of $N_\mathrm{hfc}$ hyperfine components of the $\mathrm{N_2H^+}$ emission is computed with 
\begin{equation} %
\tau_{\nu, i}^\mathrm{line} = \sum_{j=1}^{N_\mathrm{hfc}} \tau_{i,j}^\mathrm{line} (\nu) = \sum_{j=1}^{N_\mathrm{hfc}} \frac{c^2}{8 \pi \nu^2} \, \frac{N_i}{Q(T_g)} \, g_{u,j} \, A_{ul,j} \, e^{-E_{u,j}/kT_g} \, (e^{h \nu_{0,j}/k T_g} - 1) \, \phi_{i,j}(\nu),
\end{equation} %
where the subscript denotes the $j$-th hyperfine component, $\nu_{0,j}$ the rest frequency, $g_{u,j}$ the statistical weight of the upper level, $A_{ul,j}$ is the spontaneous emission rate of the transition, and $E_{u,j}$ the upper level energy.   
Values of these quantities are obtained from the Splatalogue database in National Astronomical Radio Observatory (NRAO). 
For $J=1-0$ transition, there are $N_\mathrm{hfc} = 7$ hyperfine components.  
The line profile $\phi_{i,j}(\nu)$ is given by 
\begin{equation} %
\phi_{i,j}(\nu) = \sqrt{\frac{4 \ln2}{\pi}} \, \frac{c}{\nu_{0,j} \, \Delta \upsilon_i} \, \exp \left( \displaystyle -\frac{4 \ln 2 \cdot c^2}{\Delta \upsilon_i^2} \, \left[ \frac{\nu}{\nu_{0,j}}-1+\frac{\upsilon_i}{c} \right]^2 \right).  
\end{equation} %
Therefore, the optical depth of the line emission including all the velocity components in Eq.~(\ref{eq:Tb}) is 
\begin{equation} %
\tau_\nu^\mathrm{line} = \sum_{i=1}^\Upsilon \tau_{\nu,i}^\mathrm{line},
\end{equation} %
where $\Upsilon$ is the number of velocity components in the model spectrum.  
Furthermore, the observed brightness temperature, $T_b$, may be reduced by a beam filling factor, $f_b$.  
Assuming a single filling factor for all velocity components, our model spectrum is described by 
\begin{equation} %
T_b^\mathrm{mod}(\nu) = f_b \, T_b(\nu) =  f_b \, e^{-\tau_\nu^\mathrm{cont}} \, (1-e^{-\tau_\nu^\mathrm{line}}) [J_\nu(T_g) - J_\nu(T_\mathrm{bg})].  \label{eq:Tbmod}
\end{equation} %
Note that the beam filling factor is coupled with the attenuation caused by the continuum optical depth, i.e. $e^{-\tau_\nu^\mathrm{cont}}$, in our model fitting algorithm.    

Given $\Upsilon$ velocity components along one line of sight (one pixel), we optimize the model spectrum with $n_\mathrm{par} = (1+3 \Upsilon)$ parameters for minimization of the reduced $\chi^2$ value, $\overline{\chi^2}$, using the Levenberg-Marquardt method.    
The reduced $\overline{\chi^2}$ value is normalized to the degrees of freedom, $n_\mathrm{dof}$, 
\begin{equation} %
\overline{\chi^2} \equiv \frac{\chi^2}{n_\mathrm{dof}} = \frac{\chi^2}{n_\mathrm{data} - n_\mathrm{par}}, 
\end{equation} %
where $n_\mathrm{data}$ is the number of data points and $n_\mathrm{par} = 1 + 3\Upsilon$ is the number of fitted parameters.  
For our image cubes, a maximum of four velocity components, $\Upsilon_\mathrm{max} = 4$, is sufficient to produce reasonable fits.   
To avoid underestimating emission of very narrow line width, refinement of each channel into eleven uniformly divided sub-channels in frequency is performed.  
In each frequency channel, the mean value of $T_b^\mathrm{mod}$ in all the sub-channels is used to compare with the observed value.   


Initial guess of velocity components are identified from the cross-correlation function between the observed spectrum and a template spectrum with a narrow line width of $0.1 \; \mathrm{km \, s^{-1}}$.  
For spectra with many velocity components, the cross-correlation function may not always deliver the best guess so we allow a maximum of six velocity components to serve the initial selection set.    
To determine how many velocity components are needed to fit an observed spectrum, an optimizer simply scans through all the combinations made out of the six most probable velocity components. 
The combinations of $\Upsilon$ selection out of six components are given by $\prod_{k=7-\Upsilon}^6 k/\prod_{k=1}^\Upsilon k$.  
Therefore, the maximum number of all the available combinations for an initial set of six will be 56.  
Each combination of initial guess is optimized for a solution, and the corresponding reduced $\chi^2$ value, $\overline{\chi^2}$ is computed.   
Only pixels with more than 9 channels above $3\sigma$ level are processed.     
A solution with any velocity component of spectral peak lower than $2\sigma$ is excluded from the final selection. 
We also require separation between any two velocity components to be larger than 2 channels, i.e. $0.4 \; \mathrm{km \, s^{-1}}$, to avoid excessive over-modeling.  
The solution that gives the minimum value of $\overline{\chi^2}$ among all the selected combinations is used to represent the kinematics of the working pixel.  
Note that only bright spectra of multiple peaks, such as pixels in hubs, are actually processed for all the 56 combinations of initial guess.  
After processing the entire cube, we further reject components of spectral peak lower than $3.5 \sigma$ to avoid poorly constrained components.  
This rejection is similar to the criterion in previous studies \citep[e.g.][]{Kirk:2013gq,Hacar:2018eh}.    
Although this methodology requires more computation time, it does not require visual inspection in intermediate steps and likely produces a uniform, less biased interpretation of a large data cube.  
 
\begin{figure}[h!] %
\epsscale{1.2} %
\plotone{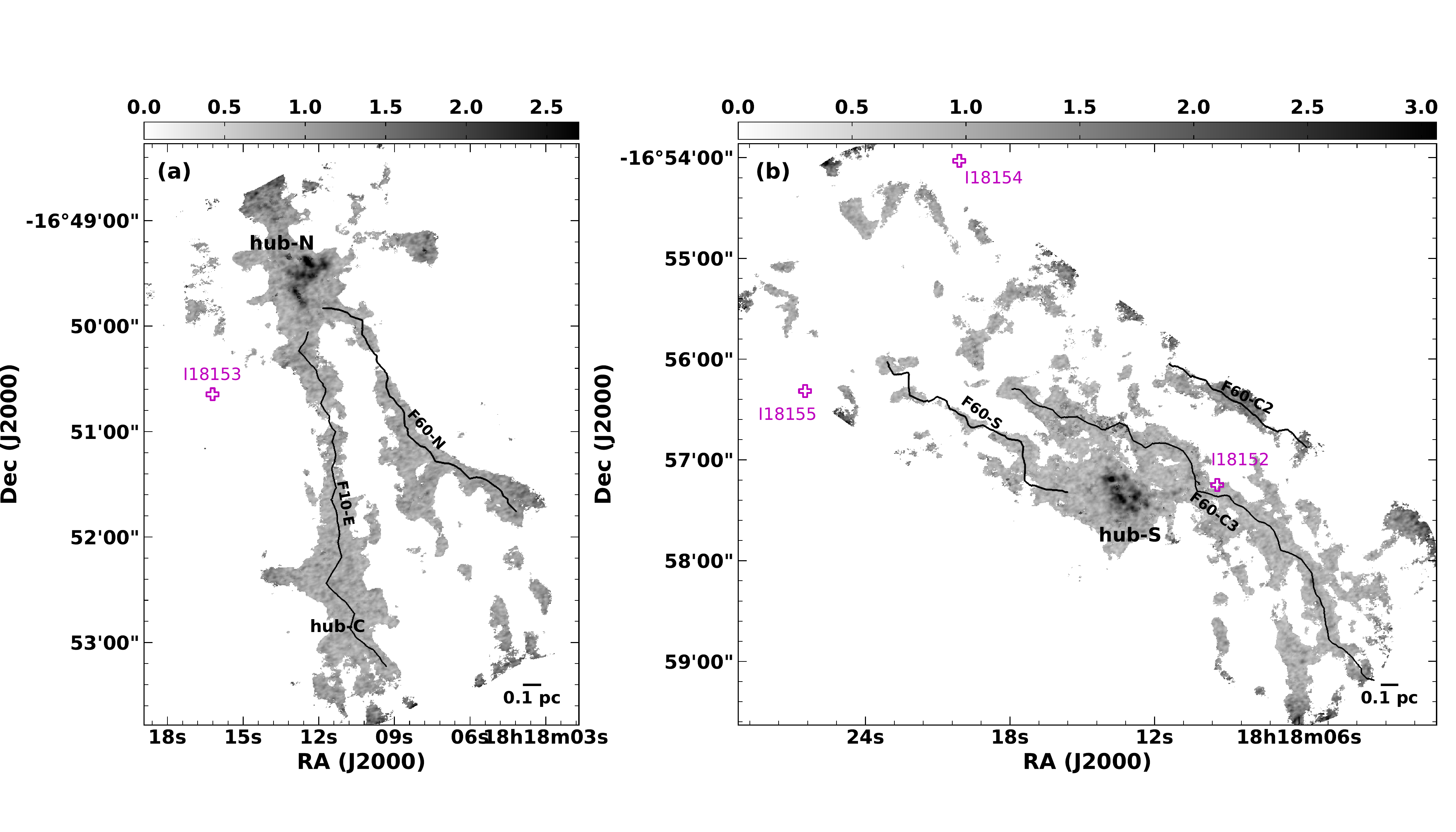} %
\caption{(a) The reduced $\chi^2$, $\overline{\chi^2}$, distribution of 149,410 pixels in Field-N.  (b) Same distribution of 199,336 pixels in Field-S.  \label{fig:rcs_maps}} %
\end{figure}

\begin{figure}[h!] %
\plotone{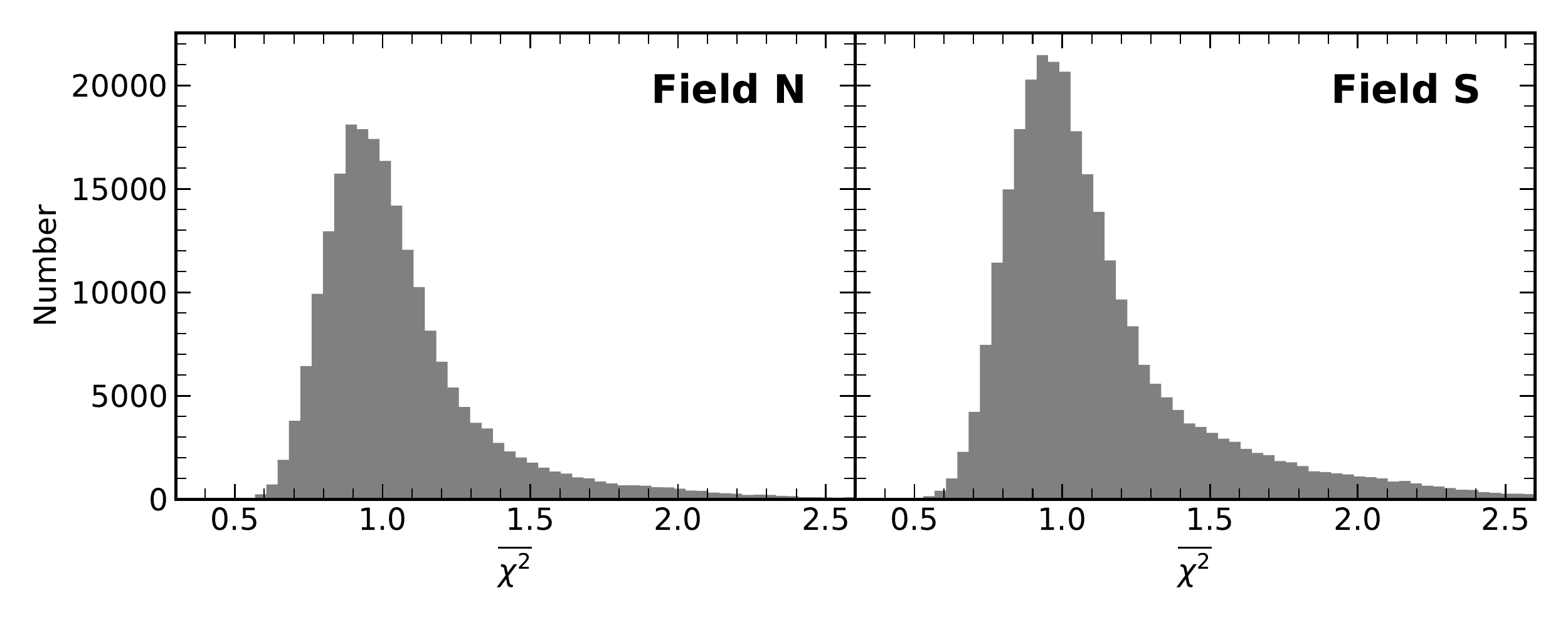} %
\caption{Histogram of $\overline{\chi^2}$ in Field-N (left) and Field-S (right).  The $\overline{\chi^2}$ distribution peaks around $\sim 0.95$ with a weak tail towards large value, which occurs in the hubs and the margin of the fields.  \label{fig:rcs_hist}} %
\end{figure}

For IRDC~G14.2, we have successfully derived the kinematics for 149,410 and 199,336 pixels in Field-N and Field-S, respectively.  
Figure~\ref{fig:rcs_maps}  shows the spatial distributions of $\overline{\chi^2}$ rendered from our fitting algorithm, and Figure~\ref{fig:rcs_hist} shows the corresponding histograms.  
Over all, pixels in the central regions of the hubs tend to have the largest $\overline{\chi^2}$ values.  
This is mostly caused by our over-simplified assumptions leading to Eq.~(\ref{eq:Tbmod}).  
For example, a temperature gradient is likely to occur in these internally heated hubs with multiple embedded YSOs. 
Besides, the beam filling factor, $f_b$, may not necessarily be a constant for all the velocity components along line of sight.   

\bibliography{myrefs} %



\end{document}